\documentclass[acmsmall,screen]{acmart}

\setcopyright{none} 
\renewcommand\footnotetextcopyrightpermission[1]{} 

\usepackage{amsmath,amsfonts}
\usepackage{verbatim}
\usepackage{graphicx}
\usepackage{textcomp}
\usepackage{xcolor}
\usepackage[utf8]{inputenc}
\usepackage{algpseudocode}
\usepackage[linesnumbered,lined,ruled,commentsnumbered]{algorithm2e}
\usepackage{xspace}
\usepackage{listings}
\usepackage{color}
\usepackage{multirow}
\usepackage{dblfloatfix}
\usepackage{lscape}
\usepackage{xparse}
\usepackage{placeins}
\usepackage{caption}
\usepackage{subcaption}
\usepackage{enumitem}
\usepackage{array, makecell}
\usepackage{listings, xcolor, soul}
\usepackage{wrapfig}
\usepackage{colortbl}
\usepackage{tabularx}
\usepackage{courier}
\usepackage{ragged2e}
\usepackage{amsthm}
\usepackage{hyperref}
\usepackage{comment}
\usepackage{pdflscape}
\usepackage{afterpage}
\usepackage{fontawesome}
\usepackage{pifont}
\usepackage{lscape}
\usepackage{enumitem}
\usepackage{booktabs}
\usepackage[most]{tcolorbox}
\usepackage{siunitx}
\usepackage{etoolbox}

\definecolor{meta-prompt}{HTML}{2D7D74}  
\definecolor{llm-response}{HTML}{2E7D32} 
\definecolor{light-gray}{gray}{0.95}     

\newcommand{\cmark}{\ding{51}}
\newcommand{\xmark}{\ding{55}}
\setlist[itemize]{noitemsep, topsep=4pt}
\definecolor{lightgray}{HTML}{fafafa}
\definecolor{darkgray}{rgb}{.55,.55,.55}
\definecolor{darkblue}{HTML}{0066cc}
\definecolor{brickred}{HTML}{b04f4f}
\definecolor{purple}{rgb}{0.65, 0.12, 0.82}
\definecolor{diffadd}{HTML}{00b359} 
\definecolor{diffrmbg}{HTML}{ffebe9}
\definecolor{diffaddbg}{HTML}{e6ffeb}
\definecolor{diffremove}{HTML}{de4f54}
\definecolor{carrotorange}{rgb}{0.8, 0.33, 0.0}
\definecolor{highlight}{HTML}{fefbc2}
\definecolor{bluegray}{HTML}{3182bd}
\definecolor{lightred}{HTML}{3182bd}

\lstdefinelanguage{JavaScript}{
  keywords={typeof, new, true, false, catch, function, return, null, catch, switch, var, const, let, extends, if, in, while, do, else, case, break, async, await, of},
  keywordstyle=\color{darkblue}\bfseries,
  ndkeywords={class, export, boolean, throw, implements, import, this, setTimeout},
  ndkeywordstyle=\color{brickred}\bfseries,
  identifierstyle=\color{black},
  sensitive=false,
  comment=[l]{//},
  morecomment=[f][\color{diffadd}\bfseries]{+\ },
  morecomment=[s]{/*}{*/},
  morecomment=[f][\color{diffremove}\bfseries]{- },
  commentstyle=\color{violet}\ttfamily,
  stringstyle=\color{carrotorange}\ttfamily,
  morestring=[b]',
  morestring=[b]"
}

\lstdefinelanguage{Python}{
  keywords={typeof, new, true, false, catch, function, return, null, catch, switch, var, const, let, extends, if, in, while, do, else, case, break, async, await, of, from, import, class, def},
  keywordstyle=\color{darkblue}\bfseries,
  ndkeywords={class, export, boolean, throw, implements, import, this, setTimeout, self, __init__},
  ndkeywordstyle=\color{brickred}\bfseries,
  identifierstyle=\color{black},
  sensitive=false,
  comment=[l]{//},
  morecomment=[f][\color{diffadd}\bfseries]{+\ },
  morecomment=[s]{/*}{*/},
  morecomment=[f][\color{diffremove}\bfseries]{- },
  commentstyle=\color{violet}\ttfamily,
  stringstyle=\color{carrotorange}\ttfamily,
  morestring=[b]',
  morestring=[b]"
}

\lstdefinelanguage{Dockerfile}{
  keywords={FROM, RUN, CMD, LABEL, MAINTAINER, EXPOSE, ENV, ADD, COPY, ENTRYPOINT, VOLUME, USER, WORKDIR, ARG, ONBUILD, STOPSIGNAL, HEALTHCHECK, SHELL},
  keywordstyle=\color{darkblue}\bfseries,
  ndkeywords={--from, --chown, --mount},
  ndkeywordstyle=\color{brickred}\bfseries,
  identifierstyle=\color{black},
  sensitive=true,
  comment=[l]{\#},
  morecomment=[f][\color{diffadd}\bfseries]{+\ },
  morecomment=[f][\color{diffremove}\bfseries]{- },
  commentstyle=\color{darkgray}\ttfamily,
  stringstyle=\color{violet}\ttfamily,
  morestring=[b]',
  morestring=[b]"
}

\definecolor{highlight}{HTML}{FFF8CC}
\definecolor{diffadd}{RGB}{0,128,0}
\definecolor{diffdel}{RGB}{180,0,0}
\definecolor{diffmeta}{RGB}{0,0,160}

\lstdefinelanguage{Diff}{
  sensitive=true,
  morecomment=[l][\color{diffmeta}]{@@},        
  morecomment=[l][\color{diffmeta}]{diff --},   
  morecomment=[l][\color{diffdel}]{-},          
  morecomment=[l][\color{diffadd}]{+},          
}

\lstdefinestyle{diff}{
  language=Diff,
  basicstyle=\ttfamily\small,
  showstringspaces=false,
  breaklines=true,
  columns=fullflexible,
  keepspaces=true,
  aboveskip=0.8\baselineskip,
  belowskip=0.8\baselineskip,
}

\theoremstyle{definition}

\newcommand{\header}[1]{\par\smallskip\noindent\textbf{#1.}}
\newcommand{\result}[1]{\par\noindent\textit{\underline{Results:}} #1}

\def\BibTeX{{\rm B\kern-.05em{\sc i\kern-.025em b}\kern-.08em
    T\kern-.1667em\lower.7ex\hbox{E}\kern-.125emX}}
    
\newboolean{showcomments}
\setboolean{showcomments}{false}
\ifthenelse{\boolean{showcomments}}
{
	\definecolor{myyellow}{RGB}{255, 228, 26}
	\definecolor{myblue}{RGB}{50, 50, 220}
	\newcommand{\nb}[2]{
		{\sf
			\fcolorbox{myyellow}{yellow}{\scriptsize\textbf{#1}}%
			$\blacktriangleright$%
			{\color{myblue}\fontsize{7pt}{8pt}\selectfont\textbf{#2}}%
		}
	}
}
{
	\newcommand{\nb}[2]{}
}

\newenvironment{change}{\begingroup\color{black}\ignorespaces}%
                       {\endgroup\ignorespacesafterend}                       

\newcommand{\changeinline}[1]{{\color{black}#1}}

\newcommand{\toolname}{\textsc{\mbox{Maple}}\xspace}
\newcommand{\toolnamefullname}{\textsc{Model Context Protocol for Automated Lightweight Repository Context Extraction}\xspace}
\newcommand{\birch}{\textsc{Birch}\xspace}
\newcommand{\hunkdefectsforj}{\textsc{Hunk4J}\xspace}

\newcommand{\hunkswe}{\textsc{HunkSWE}\xspace}

\newcommand{\benchmarkdataset}{\textsc{PolyHunk}\xspace}

\newcommand{\code}[1]{{\smaller\ttfamily\texttt{#1}}}    

\newcommand{\claudecode}{\textsc{Claude Code}\xspace}
\newcommand{\codex}{\textsc{Codex}\xspace}
\newcommand{\geminicli}{\textsc{Gemini-cli}\xspace}
\newcommand{\qwencode}{\textsc{Qwen Code}\xspace}

\newcommand{\sonnet}{\textsc{Sonnet-4.5}\xspace}
\newcommand{\gpt}{\textsc{Gpt-5}\xspace}
\newcommand{\geminimodel}{\textsc{gemini-2.5-flash}\xspace}
\newcommand{\qwencodermodel}{\textsc{qwen3-coder-flash}\xspace}

\AtBeginDocument{%
  \providecommand\BibTeX{{%
    Bib\TeX}}}

\setcopyright{acmlicensed}

\acmJournal{TOSEM}

\begin{document}

\title{%
Beyond Accuracy: Behavioral Dynamics of Agentic Multi-Hunk Repair
}

\author{Noor Nashid}
\orcid{0000-0002-9889-2312}
\affiliation{%
  \institution{The University of British Columbia}
  \city{Vancouver}
  \country{Canada}}
\email{nashid@ece.ubc.ca}

\author{Daniel Ding}
\orcid{0009-0007-7543-3926}
\affiliation{%
  \institution{The University of British Columbia}
  \city{Vancouver}
  \country{Canada}}
\email{dyxd2003@ece.ubc.ca}

\author{Keheliya Gallaba}
\orcid{0000-0002-5880-5114}
\affiliation{%
  \institution{Queen’s University}
  \city{Kingston}
  \country{Canada}}
\email{gallabak@sigsoft.org}

\author{Ahmed E. Hassan}
\orcid{0000-0001-7749-5513}
\affiliation{%
  \institution{Queen’s University}
  \city{Kingston}
  \country{Canada}}
\email{ahmed@cs.queensu.ca}

\author{Ali Mesbah}
\orcid{0000-0002-8783-7967}
\affiliation{%
  \institution{The University of British Columbia}
  \city{Vancouver}
  \country{Canada}}
\email{amesbah@ece.ubc.ca}

\begin{abstract}
Automated program repair has traditionally focused on single-hunk defects, overlooking multi-hunk bugs that are prevalent in real-world systems. 
Repairing these bugs requires coordinated edits across multiple, disjoint code regions, posing substantially greater challenges. 
We present the first systematic study of LLM-driven coding agents (\claudecode, \codex, \geminicli, and \qwencode) on this task. 
We evaluate these four state-of-the-art agents on \changeinline{404} multi-hunk bugs from the \changeinline{\benchmarkdataset} dataset, yielding \changeinline{1,616} repair trajectories for large-scale behavioral analysis. We employ fine-grained metrics to assess localization, repair accuracy, regression behavior, and operational dynamics across agents.
We find that localization capability varies substantially, with \codex achieving the highest success rate (\changeinline{75.3\%}) and \qwencode the lowest (\changeinline{40.4\%}).
Repair accuracy also differs widely, ranging from \changeinline{26.98\%} (\qwencode) to \changeinline{92.82\%} (\claudecode), and consistently declines with increasing bug dispersion and complexity (hunk divergence and spatial proximity).
High-performing agents (\claudecode and \codex) demonstrate superior semantic consistency, achieving positive average regression reduction, whereas lower-performing agents often introduce new test failures.
Notably, agents do not \emph{fail fast}; failed repairs consume substantially more resources (\changeinline{33\%–440\% more input tokens}) and require longer execution time (\changeinline{35\%–330\%}).
Additionally, we developed \toolname to provide agents with repository-level context.
Empirical results show that \toolname improves repair accuracy of \geminicli by \changeinline{$\sim$21\%} through enhanced localization.
By analyzing fine-grained metrics and trajectory-level analysis, this study moves beyond accuracy to explain how coding agents localize, reason, and act during multi-hunk repair.
Our findings underscore the impact of bug divergence and spatial proximity on multi-hunk repair success for coding agents.
\end{abstract}

%
%





\maketitle

\section{Introduction}

Automated program repair (APR) has progressed substantially in recent years, yet most existing techniques remain limited to \emph{single-hunk} bugs—defects that can be corrected by modifying a single, localized code region~\cite{hoppity, glance, katana, reptory, rewardrepair-icse-2022, recoder-fse-2021, li:dlfix:icse20, xuan:nopol:tse16, jiang:simfix:issta18, zhu:tare:icse23, yang:transplantfix:ase22}. In contrast, \emph{multi-hunk} bugs, which require coordinated edits across multiple, disjoint code regions, are prevalent in real-world software systems~\cite{hercules, birch}, yet have received limited attention~\cite{mechtaev:angelix-multi-hunk:icse16, yuan:arjae:19, wong:varfix:fse21, dear, iter}. Repairing such bugs poses significantly greater challenges, as it requires not only identifying all defective locations but also reasoning about their interdependencies and synthesizing coherent edits across the codebase.

Multi-hunk bugs differ greatly in structure and complexity, spanning edits within a single function to changes across multiple files or even packages. Our recent work~\cite{birch} characterizes this variability along two dimensions: \emph{hunk divergence}, measuring lexical and structural differences among edits, and \emph{spatial proximity}, capturing their dispersion across the codebase. Our empirical study across six large language models (LLMs) reveals that multi-hunk repair accuracy declines consistently with increasing hunk divergence and spatial dispersion~\cite{birch}.

Recent advances in AI have enabled the emergence of \emph{coding agents}, LLM-driven agents that can iteratively interact with software projects through actions such as searching, editing, building, and testing~\cite{claude-code, openai-codex, gemini-cli, qwen-code}. Unlike LLM prompting~\cite{cedar}, these agents operate autonomously over multiple steps, making decisions based on intermediate feedback from compilers, test results, and source code changes. However, understanding how these agents behave during complex software engineering tasks remains difficult. Their decision-making process is often \emph{non-deterministic}, meaning that identical inputs may lead to different sequences of actions due to the underlying model. They also adapt their strategies \emph{dynamically} as they observe new feedback, making their reasoning paths hard to interpret. This dynamic and unpredictable nature becomes even more pronounced in the context of \emph{multi-hunk} bugs, where successful repair requires identifying and consistently editing multiple related code regions. Simple outcome-based metrics, such as overall success or failure, fail to capture these nuanced behaviors. To advance automated repair, it is therefore important to analyze how coding agents reason, plan, and coordinate their actions when faced with complex, dispersed bug scenarios.

We present the first systematic study of coding agents for multi-hunk program repair. Our investigation addresses two complementary aspects. First, we evaluate how well current coding agents perform when repairing multi-hunk bugs of varying complexity. Second, we examine their underlying capabilities and behaviors during the repair process, which have not been systematically examined before. We analyze trajectory-level traces of ordered action sequences, such as search, edit, build, and test, together with the corresponding feedback from the environment, including compiler diagnostics and test results. 
Since each coding agent generates trajectories that differ in structure and interaction semantics, we normalize them into a unified representation, enabling a consistent analysis of behavioral patterns across agents.
To capture distinct aspects, we introduce a set of fine-grained metrics that separately assess localization accuracy, repair accuracy, regression, and tool usage patterns. In addition, we develop \toolname (\toolnamefullname), a lightweight code analysis system that provides coding agents with repository--level context. Our analysis reveals clear behavioral differences that cannot be observed through overall accuracy alone. Coding agents perform well on bugs with low dispersion but struggle as divergence increases, often failing to coordinate edits across multiple locations. To the best of our knowledge, this is the first work to systematically characterize both the accuracy and the behavioral mechanisms of coding agents on multi-hunk program repair tasks.

We make the following contributions:

\begin{enumerate}
    \item \begin{change} We construct \benchmarkdataset, a multi-language benchmark of 404 reproducible multi-hunk bugs in Java and Python. We extend the hunk divergence and spatial proximity metrics, originally developed for Java, to support Python. This unified framework enables both combined evaluation and language-agnostic analysis. \end{change}

    \item We conduct a systematic evaluation of four coding agents, \claudecode (Anthropic), \codex (OpenAI), \geminicli (Google), and \qwencode (Alibaba) on \changeinline{404} multi-hunk bugs from \changeinline{\benchmarkdataset}~\cite{repo}, analyzing \changeinline{1,616} repair trajectories. 

    \item We introduce a set of behavioral metrics that isolate distinct aspects of the repair capability of coding agents. \emph{Localization Success} measures the ability to identify buggy files; \emph{Compilation Success} quantifies whether agent-proposed changes are syntactically coherent and can be compiled without any errors; \emph{Repair Accuracy} captures the correctness of the generated patch; and \emph{Regression Reduction} quantifies whether an agent introduced fewer bugs than it fixed during the repair process. \result{We observe significant variation in effectiveness across agents. We find that localization success varies dramatically, from \changeinline{40\%} to 75\%, and final repair accuracy ranges from a low of \changeinline{27\%} to a high of 93\%. This highlights the substantial differences in the reasoning and code navigation capabilities of current agents. A deeper analysis of bugs not repaired by all agents reveals that high divergence can render even spatially localized bugs intractable.}

    \item We analyze tool-use sequences of agents to identify concrete behavioral patterns that distinguish successful repairs from failures. \result{The two primary failure modes exemplified in the coding agents are \emph{over-modification without validation} and \emph{over-exploration without action}.}
    
    \item \begin{change} We develop \toolname, a multilingual lightweight code analysis Model Context Protocol (MCP) tool that provides coding agents with repository--level context. \end{change} \result{Empirical results show that scope-driven, context retrieval assistance provided via MCP integration enhances the bug localization capabilities and thus improves the repair accuracy of weaker coding agents.}
\end{enumerate}

\section{Characterization of Coding Agents}
\label{sec:metrics-methodology}

The behavior of coding agents in multi-hunk code repair cannot be meaningfully characterized by a single success or failure outcome. Real-world software defects frequently span multiple, interdependent code regions~\cite{hercules}, creating complex and intertwined challenges~\cite{birch}. First, the agent must accurately localize several spatially distributed fault sites, each of which may require different contextual reasoning. Second, edits performed at one location can influence the correctness of others, either improving or degrading the overall repair outcome. This complexity is further compounded by the iterative nature of coding agents, which reason over intermediate feedback such as compiler diagnostics and test results and continuously adapt their strategies as they proceed. A comprehensive understanding of agent behavior, therefore, requires fine-grained, process-level analysis that extends beyond aggregate measures of repair accuracy.

To formalize this analysis, we employ a set of fine-grained metrics to characterize the behavioral and functional capabilities of coding agents during multi-hunk repair. These metrics collectively capture how agents identify, modify, and validate buggy code, offering a structured basis for evaluating both their effectiveness and operational behavior.

\subsection{Agent Execution Model}
We consider a set of multi-hunk bug instances \(\mathcal{B}\), where each bug \(b \in \mathcal{B}\) is associated with a set of ground-truth buggy hunks \(\mathcal{H}_{\mathrm{gt}}(b) = \{h_1, h_2, \dots, h_m\}\) and \(m \geq 2\). 
Let \(\mathcal{A}\) denote the set of coding agents under study, and let \(\mathcal{T} = \{\mathtt{search}, \mathtt{edit}, \mathtt{build}, \mathtt{test}, \ldots\}\) denote the action space comprising all available tool invocations. 
When an agent \(A \in \mathcal{A}\) is executed on a bug instance \(b\), it interacts with the development environment by performing a sequence of actions drawn from \(\mathcal{T}\), which collectively form its \emph{trajectory}~\cite{yao:react:iclr23, swe-agent, autocoderover}.

More formally, the \emph{trajectory} of a coding agent \(A\) for a given bug \(b\), denoted \( S_A(b) \), is the ordered sequence of actions executed by the agent during the repair process:
\begin{equation}
S_A(b) = \langle a_1, a_2, \ldots, a_n \rangle,
\label{eq:trajectory}
\end{equation}
where each \(a_i \in \mathcal{T}\) represents an atomic tool invocation from the available action space. 
Each action \(a_i\) produces an observation \(o_i\) (e.g., compiler output, test results, or runtime logs), which informs the agent’s subsequent decision \(a_{i+1}\). 
The trajectory, therefore, encodes the complete interaction loop between the agent and its environment, capturing how the agent plans, acts, and adapts based on feedback throughout the multi-hunk repair process.

At the end of execution, the agent produces a candidate patch, denoted by $P_A(b)$, which contains the final code modifications proposed for bug $b$. If the agent fails to complete execution, encounters a runtime error, or exceeds its time limit, the resulting patch is recorded as empty. The generated patch serves as the observable artifact of the agent’s repair process, and is later used to evaluate localization accuracy and fixing effectiveness. 

From the patch, we derive the set of code regions modified by the agent:
\begin{equation}
H(S_A(b)) = \{\, f \mid f \text{ is modified in } P_A(b) \,\}.
\end{equation}
If $P_A(b)$ is empty, due to execution failure, runtime error, or timeout, then $H(S_A(b))$ is also empty. This set represents the files examined by the coding agent and is used to analyze which files were actually modified relative to the ground-truth buggy regions.

We further record $\mathrm{Tests}^{\text{before}}_A(b)$ and $\mathrm{Tests}^{\text{after}}_A(b)$ as the numbers of passing tests before and after the agent’s execution on $b$, respectively. These quantities measure the overall impact of the agent’s repair attempt on program correctness. Overall, this execution model represents multi-hunk repair as a structured sequence of reasoning and tool-mediated interactions, forming the foundation for analyzing how coding agents plan, act, and adapt during complex repair processes.

\subsection{Effectiveness Metrics}

This section defines the \emph{effectiveness metrics} used to characterize the outcome of each repair attempt.  
These metrics capture the observable end states of the repair process, including whether the bug can be localized, whether generated patches compile successfully, and whether they restore program correctness without introducing new regressions. 

\subsubsection{Localization}
\label{sec:def-localization}

A coding agent cannot repair a bug without first identifying where the buggy code resides. In the multi-hunk setting, this challenge is exacerbated, as successful repair requires editing \emph{all} buggy regions within the codebase and performing coordinated modifications to maintain patch coherence. Localization capability, therefore, reflects how effectively the agent explores and covers the relevant buggy files required for such interdependent code changes across the project. Insufficient localization may lead to wasted edits or incomplete fixes, even when some modifications are correct.

We define the \emph{file localization success (LS)} metric as the extent to which the agent has interacted with all ground-truth buggy files while producing a plausible patch. If the generated patch $P_A(b)$ is empty or invalid because the agent failed to complete execution, encountered a runtime error, or produced uncompilable code, then the localization metric is set to 0. 

When a patch is generated, the LS metric is defined as:
\begin{equation}
\mathrm{LS}(S_A(b)) =
\begin{cases}
1, & \mathcal{H}_{\mathrm{gt}}(b) \subseteq H(S_A(b)), \\
0, & \text{otherwise.}
\end{cases}
\label{eq:localization-success}
\end{equation}

An LS value of~1 indicates that the agent has covered all ground-truth buggy files in its generated patch, even if additional files were also modified. This definition captures \emph{localization completeness}, emphasizing whether the agent successfully identified every relevant buggy file required for a coherent fix. The LS metric thus isolates the localization ability of the agent independently of the correctness of the final patch.

\subsubsection{Compilation}
Multi-hunk repairs require coordinated modifications across spatially separated code regions, where changes in one location may introduce syntactic dependencies on edits elsewhere. For instance, renaming a variable in one hunk necessitates consistent updates at all usage sites, while modifying a method signature requires corresponding changes at all call sites. Failure to maintain such consistency across hunks results in compilation errors, even when individual edits are locally correct.

We define the compilation success metric as:
\begin{equation}
\mathrm{Compilation}_A(b) =
\begin{cases}
1, & P_A(b) \neq \varnothing \land \text{the patch compiles without errors},\\[4pt]
0, & \text{otherwise.}
\end{cases}
\label{eq:compilation-success}
\end{equation}
A compilation value of 1 indicates the agent synthesized a syntactically coherent patch across all modified regions. This metric serves as a necessary precondition for functional correctness: patches that fail to compile cannot be tested and thus cannot achieve repair success. 

\subsubsection{Repair}

Localization alone does not ensure a correct repair. In the multi-hunk repair setting, fixing requires coordinated edits across buggy files so that the resulting patch compiles, passes all tests, and does not introduce new failures. Repair ability measures how effectively a coding agent produces coherent and correct patches that restore program correctness.

Following prior work~\cite{repairbench, birch}, repair accuracy is defined as whether the agent generated a plausible patch that passes all available tests:


\begin{equation}
  \mathrm{Accuracy}_A(b) =
  \begin{cases}
  1, & P_A(b) \neq \varnothing \ \text{and}\ \text{failed\_tests}_{\text{after}} = 0,\\[4pt]
  0, & \text{otherwise.}
  \end{cases}
  \label{eq:repair-accuracy}
\end{equation}

An accuracy value of 1 means that the agent produced a complete multi-hunk fix with no remaining test failures.

\subsubsection{Regression Reduction}
\label{subsec:regression-reduction}

While knowing whether a generated patch compiles and passes all originally failing tests is essential, such binary success indicators do not capture the broader behavioral impact of a repair. 
Two coding agents may achieve identical repair success rates yet differ substantially in how their patches affect the overall test outcomes. 
One agent may fix the intended failures without side effects, whereas another may resolve some tests but introduce new regressions. 
Binary metrics are unable to distinguish these scenarios.

To address this limitation, we introduce the metric \emph{regression reduction}, which quantifies the net change in the number of failing tests after applying a patch. 
For an agent \( A \) attempting to repair a bug \( b \), regression reduction is defined as:

\begin{equation}
\mathrm{RR}_A(b) =
\text{failed\_tests}_{\text{before}} -
\text{failed\_tests}_{\text{after}}
\label{eq:regression-reduction}
\end{equation}

Here, \(\text{failed\_tests}_{\text{before}}\) denotes the total number of failing test cases before applying the patch, and \(\text{failed\_tests}_{\text{after}}\) denotes the total number afterwards. 
Positive values (\(\text{RR}_A(b) > 0\)) indicate that the agent’s patch reduced the number of failing tests, either by fixing originally failing cases or by introducing fewer new failures than it resolved. 
A value of zero (\(\text{RR}_A(b) = 0\)) signifies no net change in the number of failing tests, implying that the patch compiled successfully but did not improve the test outcome. 
Negative values (\(\text{RR}_A(b) < 0\)) indicate that the patch increased the number of failing tests, suggesting a regression in correctness. 
When compilation fails, regression reduction is undefined and excluded from the analysis.

Regression reduction thus provides a finer-grained measure of repair impact than binary success metrics. 
It distinguishes agents that merely achieve correctness from those that progressively improve or degrade the test landscape. 
Agents with consistently positive regression reduction contribute net improvements, whereas those with negative averages introduce additional faults, requiring developers to correct both original and newly induced errors.

\subsection{Hunk Characteristics: Divergence and Proximity}
\label{subsec:hunk-characteristics}

Multi-hunk patches differ not only in the number of edits they contain but also in how those edits vary internally and how they are distributed across the codebase. 
To characterize these dimensions of repair complexity, we focus on two complementary properties introduced in our recent work~\cite{birch}: \emph{hunk divergence}, which captures internal heterogeneity among edits, and \emph{spatial proximity}, which reflects their distributional layout within the program structure. 
Together, these metrics provide a comprehensive view of how coordination effort scales with both the diversity and the dispersion of edits in a patch.

\subsubsection{Hunk Divergence}
While the number of hunks in a patch provides a coarse measure of its fragmentation, it does not fully capture the internal diversity among the edits. 
Patches with the same number of hunks can differ substantially in how those hunks vary lexically, structurally, or spatially. 
To quantify this variation, we adopt the notion of \emph{hunk divergence}~\cite{birch}, which characterizes the heterogeneity and coordination complexity within a multi-hunk patch.

The \emph{hunk divergence} of a multi-hunk patch \( P \) measures the degree of variation among its constituent hunks in terms of lexical, structural, and spatial dissimilarity. 
Given a patch \( P = \{h_1, h_2, \dots, h_n\} \), its divergence is defined as:
\begin{equation}
\mathrm{Div}(P) =
\ln(n) \cdot 
\left(
\frac{2}{n(n - 1)}
\sum_{1 \leq i < j \leq n}
\mathrm{Div}(h_i, h_j)
\right)
\label{eq:hunk-divergence}
\end{equation}
where $\text{Div}(h_i, h_j)$ denotes the pairwise dissimilarity between hunks \(h_i\) and \(h_j\). It quantifies how different two hunks within a patch are from each other. 
It combines three complementary dimensions: lexical, structural, and file-level separation. 
The lexical component measures the dissimilarity of code changes in terms of their textual tokens; the structural component captures differences in their abstract syntax tree (AST) representations; and the file-level term accounts for whether the edits occur within the same file or across different files. 
A context-sensitive weighting factor amplifies the file-level contribution when hunks belong to separate files, reflecting the higher coordination effort required in distributed repairs. 
The resulting score ranges from 0 to 1, where higher values indicate greater divergence between edits. This metric jointly captures the internal heterogeneity of edits and the coordination complexity induced by the number of hunks in a patch.  The overall hunk divergence score satisfies \( \text{Div}(P) \in [0, \ln(n)] \).

\subsubsection{Spatial Proximity}
\label{subsec:spatial-proximity}
In addition to internal diversity among edits, multi-hunk patches vary in how their modifications are distributed throughout the codebase. 
\emph{Spatial proximity}~\cite{birch} characterizes this distribution by describing the degree to which individual hunks are clustered within, or dispersed across, program components such as methods, classes, files, and packages. 
Whereas hunk divergence measures how edits differ, spatial proximity reflects where they occur within the program’s structural hierarchy.

Spatial proximity is represented categorically rather than as a continuous distance metric. 
Each multi-hunk patch is assigned to one of five dispersion categories, \emph{Nucleus}, \emph{Cluster}, \emph{Orbit}, \emph{Sprawl}, or \emph{Fragment}, that collectively represent increasing levels of edit dispersion.
\textit{Nucleus} patches contain hunks co-located within the same method or class; \textit{Cluster} patches extend across nearby functions or classes within a single file; \textit{Orbit} patches span multiple files within a module; \textit{Sprawl} patches affect distant files or packages; and \textit{Fragment} patches are highly scattered across the repository. 
This hierarchical representation offers an interpretable view of edit dispersion that aligns with developers’ reasoning about code locality and coordination effort. 

Hunk divergence and spatial proximity together serve as key indicators of repair difficulty. 
Divergence captures how heterogeneous the edits are, while proximity captures how far apart they are distributed. 
Patches with high divergence and low proximity tend to involve varied yet localized changes, whereas those with both high divergence and high dispersion require extensive coordination across multiple interdependent code regions. 
These complementary perspectives allow us to characterize the structural and spatial complexity that coding agents must overcome during multi-hunk repair.

\subsection{Operational Dynamics of Coding Agents}
Beyond outcome-oriented measures such as repair accuracy and regression reduction, we examine a complementary set of behavioral metrics that characterize operational dynamics of coding agents during the repair process.  These metrics capture the efficiency, runtime dynamics, and interaction patterns underlying each repair attempt.

\subsubsection{Token Consumption}
\label{subsec:token-consumption}

We quantify \emph{token consumption} to measure the number of language model tokens exchanged between the agent and the API throughout the repair trajectory. 
For an agent \(A\) repairing bug \(b\), we distinguish two complementary components: \emph{input tokens} transmitted to the model and \emph{output tokens} generated by the model across all invocations. 
Let \(K_b\) denote the number of model calls made during the repair process. 
The total counts are defined as:

\begin{equation}
\mathrm{InTok}_A(b) = 
\sum_{k=1}^{K_b} \mathrm{tokens\_sent}_A^{(k)}, 
\quad
\mathrm{OutTok}_A(b) = 
\sum_{k=1}^{K_b} \mathrm{tokens\_generated}_A^{(k)}.
\label{eq:token-consumption}
\end{equation}

The overall token consumption is then given by:
\begin{equation}
\mathrm{TC}_A(b) =
\mathrm{InTok}_A(b) +
\mathrm{OutTok}_A(b)
\label{eq:total-token-consumption}
\end{equation}

Input tokens capture all prompt content, tool calls, and contextual information provided to the model, while output tokens correspond to the model’s generated responses. 
These quantities together characterize the operational footprint of a repair attempt, serving as a proxy for the computational efficiency and resource utilization of coding agents.

\subsubsection{Runtime}
\label{subsec:runtime}

Runtime measures the wall-clock time required for an agent to complete a repair task, from task initialization to termination. This metric captures the end-to-end execution duration encompassing all agent activities, including model inference calls, tool invocations, code modifications, compilation checks, and test executions throughout the repair process.

For a bug $b$ and agent $A$, runtime is defined as:

\begin{equation}
\mathrm{Runtime}_A(b) = t_{\text{end}} - t_{\text{start}}
\label{eq:runtime}
\end{equation}
\noindent
where $t_{\text{start}}$ is the timestamp when the agent receives the repair task, and $t_{\text{end}}$ is the timestamp when the agent terminates, either through successful repair, timeout, or explicit failure.

Runtime serves as a practical efficiency metric complementary to token consumption. While token consumption measures computational cost in terms of model inference, runtime reflects the actual elapsed time required for repair completion, which is critical for assessing agent responsiveness in practical scenarios. 

\subsubsection{Agentic Tool Utilization}
We characterize coding agent behavior through trajectory analysis, which models the complete sequence of actions performed during repair, including tool invocations such as \code{search}, \code{edit}, \code{build}, and \code{test}. These trajectories capture how agents localize buggy regions, decide when and where to apply edits, and adapt to feedback. By examining the frequency and distribution of tool usage within these trajectories, we identify recurring behavioral patterns that reveal how agents allocate effort and coordinate progress across multiple bug locations.

For a trajectory $S_A(b)$ (Equation~\ref{eq:trajectory}), we record the number of calls to each action type $a \in \mathcal{A}$:
\begin{equation}
\mathrm{calls}(a) = 
\bigl|\{\, i \mid a_i = a \,\}\bigr|
\label{eq:tool-calls}
\end{equation}

\noindent
The relative frequency of each tool is then defined as:
\begin{equation}
U_a =
\frac{\mathrm{calls}(a)}{\sum_{a' \in \mathcal{A}} \mathrm{calls}(a')}
\label{eq:tool-utilization}
\end{equation}

We leverage these metrics to understand how much each tool contributes to the agent’s repair process. Moreover, we investigate whether the agent distributes effort across multiple tools or relies on a limited subset of actions.

\subsubsection{Tool Sequencing Pattern} 
Developers typically follow structured tool-use sequences while debugging. For instance, a common workflow involves diagnosing the problem, applying edits, running tests to observe behavior, and refining the changes based on feedback. We examine whether coding agents exhibit similar sequencing patterns in their workflow.

Given that trajectories capture the temporal order of the agent’s reasoning and tool usage during repair, we analyze each trajectory $S_A(b)$ using a fixed window size $w$ to extract local subsequences of actions:
\begin{equation}
p_i = 
\langle a_i, a_{i+1}, \ldots, a_{i+w} \rangle,
\quad
1 \leq i \leq n - w
\label{eq:tool-sequence}
\end{equation}

For each agent, we aggregate all subsequences across its trajectories and identify the top-$n$ most frequent patterns. These top-$n$ sequences capture the agent’s characteristic repair workflows during multi-hunk fixing. 

By examining these top-$n$ sequences, we aim to assess whether coding agents exhibit coherent diagnostic patterns, such as test-driven development (edit–test–refine) loops similar to those of human developers, or follow distinct workflows that reflect different reasoning strategies. This analysis seeks to understand how agents coordinate actions, whether they leverage test feedback to guide refinement, and to what extent their operational structure aligns with systematic debugging practices. Consistent sequencing patterns would indicate that the agent organizes its repair process into structured operational cycles, whereas fragmented or highly variable patterns may reveal reactive or uncoordinated reasoning during multi-hunk repair.

\section{\toolname}
\label{sec:maple}

Effective debugging and modification depend on structured navigation and context awareness. To repair multi-hunk bugs, coding agents must locate and reason over multiple, distributed code regions. However, large repositories exceed the model's context window, preventing agents from loading entire projects at once. This limitation makes efficient repository exploration essential for scaling automated repair.

We present a systematic mechanism that equips agents with developer-like navigation capabilities through repository-level context extraction. Given a codebase, our method parses and indexes its structure, extracting type definitions (classes, interfaces, enums) and method declarations. It provides multi-granularity access—global, file-level, and class-level—enabling agents to explore codebases incrementally rather than monolithically. Agents can query repository structure, retrieve class skeletons, or fetch specific methods and code snippets on demand.

\begin{change}
\header{Background: Model Context Protocol}
The Model Context Protocol (MCP)~\cite{mcp-spec} is an open standard that defines how LLM-based agents interface with external tools. In this setup, a coding agent acts as the client, while an MCP server provides access to tools; communication occurs via JSON-RPC messages over standard input/output or HTTP. A server exposes a collection of tools, each defined by a name, a brief description, and a JSON Schema that specifies its arguments. At the start of a session, the client retrieves the list of available tools along with their schemas. To invoke a tool, the client sends the tool name and corresponding arguments, and then receives a response. Because the protocol is agent-agnostic, a single server can be used by any MCP-compatible client, including \qwencode, \geminicli, \codex, and \claudecode, without requiring agent-specific customization.
\end{change}

\header{Design}
Instead of implementing ad-hoc tools for each agent, we adopt the Model Context Protocol (MCP)~\cite{mcp-spec}.
Our approach builds on established approaches for repository-level context extraction~\cite{contextual-api-completion:arxiv24, autocoderover, issue2test}.
Our MCP server, \toolname, implements nine commands organized by scope and granularity, each prefixed with \code{maple\_} to prevent naming conflicts with other MCP endpoints. Table~\ref{tab:mcp-tools} summarizes these capabilities.

\begin{table*}[t]
\caption{\toolname MCP tools}
\label{tab:mcp-tools}
\centering
\scriptsize
\begin{tabular}{@{}lllll@{}}
\toprule
\textbf{Scope} & \textbf{Tool} & \textbf{Input} & \textbf{Returns} & \textbf{Purpose} \\
\midrule
\textbf{Class} & \code{maple\_find\_method\_in\_class} & method, class & Method source code & Locate method within given class \\
\midrule
\multirow{4}{*}{\textbf{File}} & \code{maple\_find\_class\_in\_file} & class, file & Class declaration & Locate class within specific file \\
& \code{maple\_find\_method\_in\_file} & method, file & Method source code & Locate method within specific file \\
& \code{maple\_find\_code\_in\_file} & snippet, file & Code with context & Search code in specific file \\
& \code{maple\_extract\_class\_skeleton} & file name & Package, imports, signatures & Extract class structure \\
\midrule
\multirow{4}{*}{\textbf{Global}} & \code{maple\_find\_class} & class name & Class declaration & Locate class in codebase \\
& \code{maple\_find\_method} & method name & Method source code & Locate method in codebase \\
& \code{maple\_find\_code} & code snippet & Code with context & Search code in codebase \\
& \code{maple\_repo\_structure} & project path & Directory tree & View repository structure \\
\bottomrule
\end{tabular}
\end{table*}

To support these operations efficiently, we construct structural indices based on Abstract Syntax Trees (ASTs), which precisely capture program elements such as class declarations, method definitions, and type hierarchies. Unlike text-based search utilities (e.g., \code{grep}, \code{awk}), AST parsing distinguishes code structure from literals and comments, enabling accurate retrieval of semantic entities. For each file, we extract types and methods with their fully qualified names and locations, building hierarchical indices that map methods to their enclosing classes and classes to their source files. Unparseable files are logged but excluded from the index. These indices are generated at session initialization and persist throughout the repair process, allowing fast, context-aware lookups without repeated parsing.

\begin{change}
\toolname{} indexes the codebase of the project by recursively enumerating files from the project root. Compiled artifacts (bytecode and packaged archives) and externally installed dependencies (package-manager caches and virtual environment libraries) are excluded. These elements are not accessible through any \code{maple\_*} tool. We use ASTs as a language-agnostic representation. Most programming languages provide a parser that converts source code into an AST as the first stage of compilation. The structural elements we require, such as type and callable declarations, are consistently represented across languages. Our approach is lightweight and scalable. It requires only a single linear pass over the source code. It does not depend on build systems, execution environments, or whole-program analysis. In contrast, whole-program analyses, such as type inference, call graph construction, points-to analysis, and dataflow analysis, are computationally expensive and language-specific. These requirements limit their scalability in large repositories.
\end{change}
\section{Evaluation}
\label{sec:evaluation}

Our study addresses the following research questions:

\begin{itemize}

\item \textbf{RQ1}: How effectively can coding agents localize multi-hunk bugs? 

\item \textbf{RQ2}: What is the repair accuracy of coding agents when addressing multi-hunk bugs?

\item \textbf{RQ3}: How do bug characteristics, such as hunk divergence and spatial proximity, affect the repair success rate of coding agents?

\item \textbf{RQ4}: To what extent do coding agents introduce regressions during multi-hunk bug repair?

\item \textbf{RQ5}: How do the operational dynamics of coding agents, including token consumption, tool utilization, sequencing behavior, and runtime, reflect their internal mechanisms during multi-hunk bug repair?

\item \textbf{RQ6}: Does \toolname, our scope-driven context-assisting MCP, enhance the ability of coding agents to repair multi-hunk bugs?
\end{itemize}

\begin{change}
\subsection{Dataset}
\label{sec:dataset}
We construct \benchmarkdataset, a multi-language benchmark of reproducible multi-hunk bugs, by combining two subsets: \hunkdefectsforj, a curated collection of multi-hunk bugs from open-source Java projects, and \hunkswe, a collection of multi-hunk bugs from open-source Python projects. The two subsets are introduced and described separately, as they differ in language and construction methodology. Their combination extends the benchmark beyond a single language and supports language-specific comparison.

\header{\hunkdefectsforj} To enable a systematic evaluation of multi-hunk program repair, we use \hunkdefectsforj~\cite{birch}, a curated dataset of reproducible multi-hunk bugs with test-suite validation collected from open-source Java projects.

\hunkdefectsforj contains 372 developer-written multi-hunk bugs with patches containing multiple, distinct modification hunks. Single-file multi-hunk bugs constitute the majority of the cases, with 244 instances where all changes are contained within a single file. Among these, 140 involve two hunks, 55 contain three hunks, and 49 include four or more. Such cases represent intra-file repair scenarios that primarily demand localized reasoning. In contrast, 128 bugs extend across multiple files, reflecting more fragmented and distributed repair patterns. Of these, 68 include four or more hunks, indicating substantial dispersion of changes throughout the codebase.

By integrating test-suite-validated patches with fine-grained hunk annotations, \hunkdefectsforj enables systematic analysis of bug heterogeneity across multiple dimensions, such as lexical variation, structural complexity, file-level dispersion, and spatial distribution of edits.

\begin{change}
\header{\hunkswe}
\label{sec:hunkswe}
We introduce \hunkswe, a curated dataset of reproducible multi-hunk bugs with test-suite validation for Python projects, constructed from SWE-bench Verified~\cite{swebench}. It complements \hunkdefectsforj by covering Python projects. SWE-bench Verified is a 500-instance subset of SWE-bench in which each issue has been reviewed by professional Python developers for solvability, test correctness, and reproducibility. Each instance includes a real GitHub issue, the target buggy commit, the developer-written patch, an executable test environment, and two test sets for evaluation: \code{FAIL\_TO\_PASS}, which must pass after repair, and \code{PASS\_TO\_PASS}, which must remain passing.

We construct \hunkswe by applying two filters to SWE-bench Verified. Of the 500 instances, 220 (44\%) are multi-hunk and 280 (56\%) are single-hunk; we retain only the multi-hunk instances. We then exclude instances whose issue label is not ``Bug'', removing feature requests, documentation-only issues, and unlabeled cases. This filter is the main constraint, as most multi-hunk instances correspond to feature changes or refactoring. The resulting dataset contains 32 instances from twelve Python projects, including \code{astropy}, \code{django}, \code{sympy}, \code{matplotlib}, \code{scikit-learn}, and \code{requests}.

\end{change}

\header{\benchmarkdataset}
\label{sec:poly-hunk}
We define the union of \hunkdefectsforj and \hunkswe as \benchmarkdataset, a multi-language benchmark consisting of 404 reproducible multi-hunk bugs, including 372 Java bugs from \hunkdefectsforj and 32 Python bugs from \hunkswe. All four coding agents are evaluated on this dataset under the same conditions, using a shared prompt template and evaluation harness. Reporting results on \benchmarkdataset enables direct comparison across agents. The hunk divergence and spatial proximity metrics described in Section~\ref{sec:metrics-methodology} are applied consistently to both subsets. This allows comparisons across subsets, such as differences in divergence between fixed and unfixed bugs, and accuracy across different spatial proximity categories.

  \begin{table}[t]
  \scriptsize
  \caption{Spatial proximity classes and mean hunk divergence for multi-hunk bugs in \benchmarkdataset}
  \label{tab:proximity_class:hunk-poly}
  \centering
  \begin{tabular}{l|cc|cc|cc}
  \toprule
   & \multicolumn{2}{c|}{\textbf{\hunkdefectsforj} (372)} & \multicolumn{2}{c|}{\textbf{\hunkswe} (32)} & \multicolumn{2}{c}{\textbf{\benchmarkdataset} (404)} \\
  \textbf{Proximity Class} & \textbf{\#Bugs} & \textbf{Mean Div.} & \textbf{\#Bugs} & \textbf{Mean Div.} & \textbf{\#Bugs} & \textbf{Mean Div.} \\
  \midrule
  Nucleus  & 59  & 0.2548 & 6  & 0.0952 & 65  & 0.2400 \\
  Cluster  & 185 & 0.4280 & 16 & 0.1438 & 201 & 0.4054 \\
  Orbit    & 67  & 0.5628 & 5  & 0.4119 & 72  & 0.5523 \\
  Sprawl   & 50  & 0.6718 & 1  & 0.7929 & 51  & 0.6742 \\
  Fragment & 11  & 0.7372 & 4  & 0.8573 & 15  & 0.7692 \\
  \bottomrule
  \end{tabular}
  \end{table}

Table~\ref{tab:proximity_class:hunk-poly} groups bugs in \benchmarkdataset by spatial proximity class. The distribution is skewed toward localized fixes. \emph{Cluster} is the largest class (201 bugs), with hunks in the same file but different methods. \emph{Orbit} (72 bugs) spans multiple files within a package, and \emph{Nucleus} (65 bugs) is confined to a single method. \code{Mockito\_6}~\cite{mockito6-multi-hunk-patch} is a \emph{Cluster} example with 20 hunks in one file. \emph{Sprawl} (51 bugs) and \emph{Fragment} (15 bugs) exhibit greater spread. \code{Closure\_144}~\cite{closure144-multi-hunk-patch} is a \emph{Fragment} example with hunks across unrelated packages. Mean hunk divergence increases with spatial spread; \emph{Fragment} patches are rare but have the highest mean divergence (0.7692). The subsets differ in distribution. \hunkdefectsforj covers all five classes, mainly \emph{Cluster} (185) and \emph{Orbit} (67). \hunkswe is concentrated in \emph{Cluster} (16) and \emph{Nucleus} (6), with only one \emph{Sprawl} case.
\end{change}

\subsection{Agents}
We evaluate four state-of-the-art coding agents in our multi-hunk bug repair study: \codex\footnote{\url{https://openai.com/codex/}} (OpenAI), \geminicli\footnote{\url{https://cloud.google.com/gemini/docs/codeassist/gemini-cli}} (Google), \claudecode\footnote{\url{https://www.claude.com/product/claude-code}} (Anthropic), and \qwencode\footnote{\url{https://github.com/QwenLM/qwen-code}} (Alibaba). These agents were selected because they represent four leading industrial efforts in autonomous code generation and repair. Together, they capture diverse design philosophies and interaction paradigms across the current landscape of tool-augmented LLM systems.

\codex, developed by OpenAI, is one of the most widely adopted coding agents. \geminicli, part of Google’s Gemini Code Assist ecosystem, integrates closely with real-world developer environments through command-line and IDE interfaces. \claudecode, from Anthropic, focuses on interpretable code manipulation and task execution, enabling multi-step problem solving through structured reasoning loops. Finally, \qwencode, released by Alibaba Cloud, is an open-source multilingual code model designed for reproducible research and extensibility across programming tasks.
These four agents collectively span a broad spectrum of model architectures, tool integrations, and accessibility levels, from proprietary commercial platforms to open research frameworks.

\subsection{Procedure}


We developed a fully automated framework to evaluate each coding agent on all bugs in \benchmarkdataset. While each agent has distinct command-line interfaces, underlying models, and output formats, we execute them in a standardized, automated fashion without manual intervention.

For each bug, we execute a five-stage pipeline. First, we checkout the buggy version into an isolated workspace directory. Second, we inject bug-specific metadata (title and description from the original bug report) into a standardized prompt template (Figure~\ref{fig:prompt-template}). Third, we provide additional scripts (\code{run\_bug\_exposing\_tests.sh}, \code{run\_all\_tests\_trace.sh}) that generate complete test execution trace logs with full stack traces, enabling agents to diagnose failures more effectively. Fourth, we launch the agent in headless mode. As an example, for \qwencode, we invoke \code{qwen-code --sandbox --yolo -p "Read and execute AGENT.md" --telemetry}, where \code{--sandbox} restricts file system operations to the workspace, \code{--yolo} enables autonomous action execution without approval prompts, and \code{--telemetry} captures complete trajectory traces including tool invocations and model responses. Fifth, after agent termination, we run compile and test commands to determine patch correctness.

All agents receive identical task descriptions through a standardized prompt template. Figure~\ref{fig:prompt-template} shows the condensed structure. We add bug-specific metadata, including the title and description from the original bug report. The template specifies the repair workflow and documents the available build and test commands. For RQ6, the prompt includes additional descriptions of \toolname MCP tools for structured codebase search. The complete prompt is provided in our replication package.

\begingroup
\setlength{\abovecaptionskip}{2pt}
\begin{figure}[t]
    \centering
    \begin{tcolorbox}[
        colback=meta-prompt!10!white, colframe=meta-prompt,
        boxrule=1.5pt, rounded corners, title=\textbf{Task Description},
        boxsep=3pt, top=4pt, bottom=4pt, left=5pt, right=5pt,
        fontupper=\small
    ]
\noindent\textbf{Task:} Fix a bug in a Java project. The project contains failing test cases. Investigate and resolve the underlying defect to restore full test passage.

\vspace{2pt}
\noindent\textbf{Bug Context:}
\begin{itemize}[left=0pt, itemsep=1pt, topsep=1pt]
    \item \textbf{Title:} {\footnotesize\texttt{\{\{bug\_title\}\}}}
    \item \textbf{Description:} {\footnotesize\texttt{\{\{bug\_description\}\}}}
\end{itemize}

\vspace{2pt}
\noindent\textbf{Workflow:} Identify failing tests, diagnose root cause, apply fixes, verify correctness through comprehensive testing, and ensure no regressions.

\vspace{2pt}
\noindent\textbf{Build \& Test Commands:}
\begin{itemize}[left=0pt, itemsep=1pt, topsep=1pt]
    \item Compile: {\footnotesize\texttt{defects4j compile}}
    \item Test: {\footnotesize\texttt{defects4j test}, \texttt{./run\_bug\_exposing\_tests.sh}, \texttt{./run\_all\_tests\_trace.sh}}
\end{itemize}

\vspace{2pt}
\noindent\textbf{Maple MCP -- Structured Codebase Search:}
\begin{itemize}[left=0pt, itemsep=1pt, topsep=1pt]
    \item Class: {\footnotesize\texttt{maple\_find\_method\_in\_class}}
    \item File: {\footnotesize\texttt{maple\_find\_class\_in\_file}}, {\footnotesize\texttt{maple\_extract\_class\_skeleton}}
    \item Global: {\footnotesize\texttt{maple\_find\_class}}, {\footnotesize\texttt{maple\_repo\_structure}}
\end{itemize}

\vspace{2pt}
\noindent\textbf{Success Criteria:} Zero compilation errors, zero test failures.
    \end{tcolorbox}
    \caption{Standardized prompt template provided to all coding agents (condensed).}
    \label{fig:prompt-template}
\end{figure}
\endgroup


Each agent executes within its own isolated workspace directory containing the checked-out buggy project, eliminating cross-contamination between bug instances. Agents have unrestricted access to their workspace file system, the build and test framework, and their respective API endpoints. We access \claudecode through the Claude Max plan (\$200/month), \codex through ChatGPT Pro (\$200/month), \geminicli using \geminimodel from Google, and \qwencode using \qwencodermodel through OpenRouter\footnote{\url{https://openrouter.ai/}}. All agents operate in headless mode.

\changeinline{We conducted the evaluation on \hunkdefectsforj between September and November 2025, and on \hunkswe between April and May 2026.} At the beginning of the study, the following agent CLI versions were used: \claudecode~2.0.13, \codex~0.21.0, \geminicli~0.10.0, and \qwencode~0.0.11.
\changeinline{In this evaluation, each agent invokes a specific underlying model. \claudecode uses Anthropic's \textsc{Sonnet-4.5}, \codex uses OpenAI's \textsc{GPT-5}, \geminicli uses Google's \textsc{Gemini-2.5-Flash}, and \qwencode uses Alibaba's \textsc{qwen3-coder-flash} served through OpenRouter. We treat these agent--model pairings as fixed across all instances of \benchmarkdataset.}


Upon task completion, we first compile the patched code. Compilation failures result in immediate rejection. For compilable patches, we run the full test suite. Test outcomes are recorded in structured CSV files. We further collect trajectory logs and patch diffs from each agent.

\begin{change}
\subsection{Implementation}
\label{subsec:implementation} 
Our implementation is in Python. For hunk divergence and spatial proximity on \hunkdefectsforj{}, we reuse the implementation from \birch{}~\cite{birch}. For \hunkswe{}, we implement both metrics using the \code{ast} module.

Agent automation is implemented through per-agent Python wrappers, one for each of \claudecode{}, \codex{}, \geminicli{}, and \qwencode{}. Bugs from \hunkdefectsforj{} are repaired on the host system. Bugs from \hunkswe{} are repaired inside per-instance Docker containers built from the official SWE-bench base image. Patch evaluation for \hunkswe{} is delegated to the official SWE-bench evaluation harness, which is invoked as a subprocess.

\toolname{} is implemented as an MCP server using FastMCP\footnote{\url{https://github.com/jlowin/fastmcp}}. It uses \code{javalang} for Java codebases and the \code{ast} module for Python codebases. The server is initialized at the project root for each bug. It enumerates source files via a recursive search, excluding standard build output directories. The index is accessed through the nine \code{maple\_*} tools listed in Table~\ref{tab:mcp-tools}.
\end{change}

\begin{change}
\begin{table*}
\scriptsize
\centering
\caption{Multi-hunk bug repair with coding agents}
\label{tab:multiagent-accuracy-combined}
\begin{tabular}{@{}lrrrrr@{}}
\toprule
\textbf{Agent} & \textbf{Total} & \textbf{Localization} & \textbf{Compilation} & \textbf{Regression} & \textbf{Repair} \\
 & \textbf{Bugs} & \textbf{Success (\%)} & \textbf{Success (\%)} & \textbf{Reduction} & \textbf{Accuracy (\%)} \\
\midrule
\qwencode   & 404 & 163 (40.35\%) & 392 (97.03\%)  & -1.50 & 109 (26.98\%) \\
\geminicli  & 404 & 199 (49.26\%) & 379 (93.81\%)  & -2.10 & 174 (43.07\%) \\
\codex      & 404 & 304 (75.25\%) & 399 (98.76\%)  &  2.16 & 353 (87.38\%) \\
\claudecode & 404 & 267 (66.09\%) & 404 (100.00\%) &  2.17 & 375 (92.82\%) \\
\bottomrule
\end{tabular}
\end{table*}
\end{change}


    

\subsection{Localization Ability (RQ1)}
\label{sec:localization}

As defined in Section~\ref{sec:def-localization}, an agent successfully localizes bug $b$ if the set of files it modifies is a superset of the files modified in the developer patch. This criterion ensures that for multi-hunk bugs spanning multiple files, the agent must identify all fault locations. Localization success is necessary but not sufficient for repair success: an agent that fails to localize all buggy files cannot produce a correct fix. However, multiple semantically equivalent patches may exist for the same bug, potentially modifying different sets of files. Developer patches provide a consistent, reproducible ground truth for comparative evaluation across agents. \changeinline{While this file-level localization criterion does not account for patch equivalence, it ensures that all agents are evaluated against the same localization target.}


\begin{change}
Table~\ref{tab:multiagent-accuracy-combined} presents localization success rates across all four agents. \codex yields the highest success rate at 75.25\% (304 bugs), followed by \claudecode at 66.09\% (267 bugs), \geminicli at 49.26\% (199 bugs), and \qwencode at 40.35\% (163 bugs). The gap between the highest and lowest rates is approximately 35 percentage points, and \codex localizes nearly twice as many bugs as \qwencode. \codex and \claudecode both exceed 65\%, while \geminicli and \qwencode remain below 50\%. These results indicate that current coding agents differ substantially in their ability to navigate complex repository structures.
\end{change}


\subsection{Repair Ability (RQ2)}
\label{sec:repair}



\begin{change}
Table~\ref{tab:multiagent-accuracy-combined} presents compilation success and repair accuracy across all four agents on 404 \benchmarkdataset bugs. We first examine compilation success, then analyze repair accuracy.

\header{Compilation Success} Compilation success measures whether the generated patch produces syntactically valid code that compiles without errors. \claudecode achieves perfect compilation success at 100\% (404 bugs), indicating all generated patches compile successfully. \codex and \qwencode also achieve high compilation rates at 98.76\% (399 bugs) and 97.03\% (392 bugs) respectively. \geminicli exhibits the lowest compilation success at 93.81\% (379 bugs), with 25 bugs producing patches that fail to compile. These near-perfect compilation rates across all coding agents indicate that syntactic correctness is not a primary bottleneck for multi-hunk bug repair. Even the lowest-performing agent compiles successfully 93.81\% of the time.
\end{change}



\begin{change}
\header{Repair Accuracy} \claudecode yields the highest repair accuracy at 92.82\% (375 bugs), followed by \codex at 87.38\% (353 bugs). \geminicli achieves 43.07\% (174 bugs) and \qwencode achieves 26.98\% (109 bugs). Thus, repair accuracy varies substantially across agents, ranging from 26.98\% (\qwencode) to 92.82\% (\claudecode). While compilation success remains high across agents (93.81\%--100\%), repair accuracy varies widely. This gap indicates that syntactic correctness is necessary but insufficient for successful repair.

\claudecode and \codex demonstrate superior ability to coordinate edits across multiple interdependent locations while preserving semantic consistency. These agents, built on \sonnet and \gpt respectively, exhibit a stronger capacity to reason over distributed context and reconcile dependencies across files and hunks. In contrast, \geminicli and \qwencode frequently produce patches that compile but fail to fix bugs. The 43.07\% accuracy of \geminicli indicates that less than half of compilable patches achieve functional correctness, whereas the 26.98\% accuracy of \qwencode indicates that roughly one in four patches successfully repair bugs.
\end{change}


\begin{figure}
  \centering
  \includegraphics[width=0.40\linewidth]{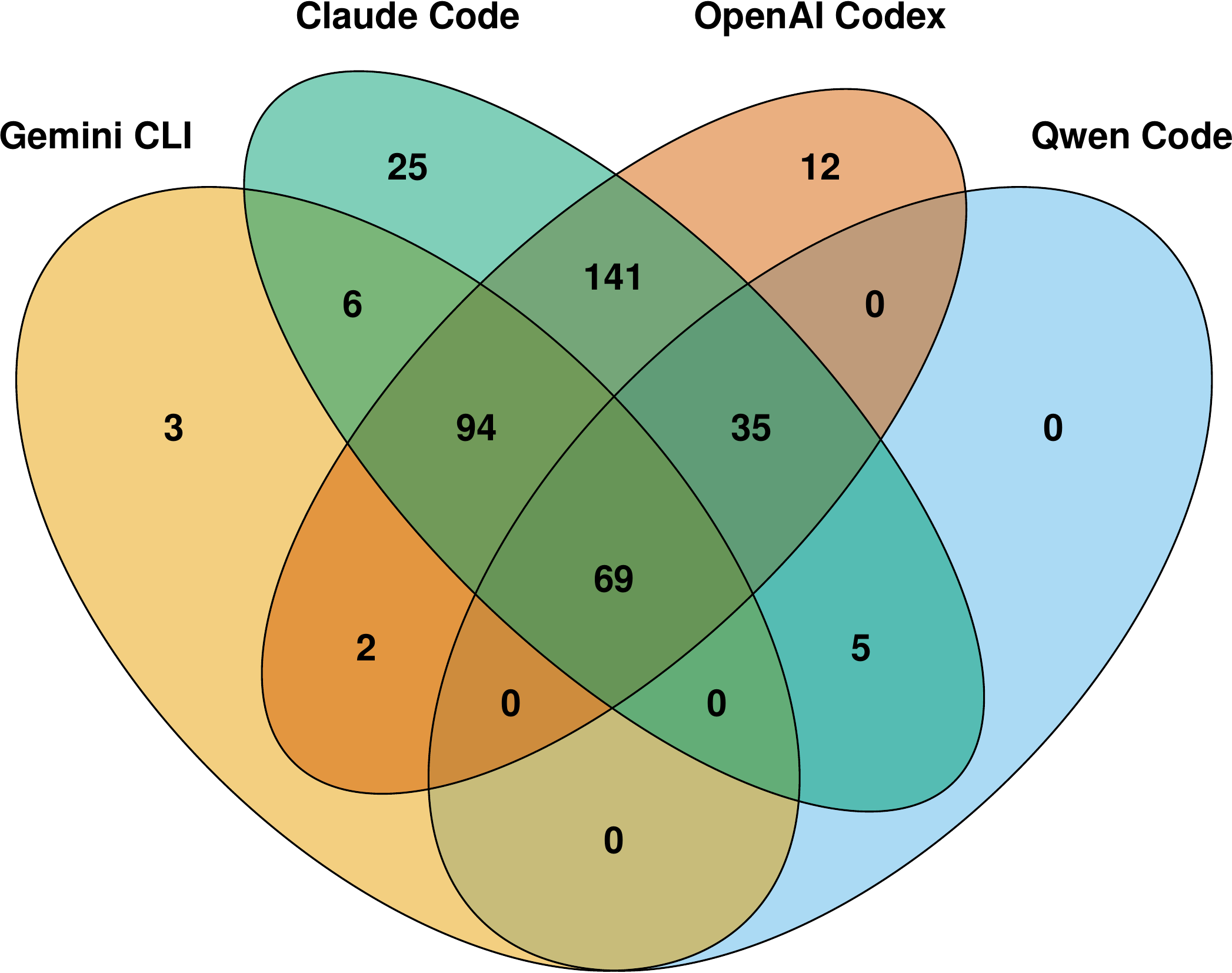}
  \caption{Overlap of fixes across coding agents}
  \label{fig:venn-llm}
\end{figure}



\header{Overlap of Agent Capability}
\begin{change} Figure~\ref{fig:venn-llm} shows the intersection of correctly repaired bugs across the four coding agents. The central overlap contains 69 bugs that all agents, \claudecode, \codex, \geminicli, and \qwencode—successfully repair, representing a shared subset of defects consistently solvable across different model architectures.
Beyond this common subset, \claudecode contributes 25 unique repairs, followed by \codex with 12 and \geminicli with 3, while \qwencode provides no unique fixes.
\end{change}

\begin{change}
The three-way intersection of \claudecode, \codex, and \geminicli covers an additional 94 bugs. A smaller three-way intersection is observed for \claudecode, \codex, and \qwencode (35 bugs). The intersection of \claudecode, \geminicli, and \qwencode (excluding \codex) contains zero bugs. The intersection of \codex, \geminicli, and \qwencode (excluding \claudecode) also contains zero bugs. Pairwise-only intersections are concentrated in a few agent combinations. \claudecode and \codex share 141 repairs that no other agent produces. The remaining pairwise-only counts are smaller: 6 bugs are repaired by only \claudecode and \geminicli, 5 by only \claudecode and \qwencode, and 2 by only \codex and \geminicli.
\end{change}
These results indicate measurable complementarity among the agents: high-accuracy agents such as \claudecode and \codex not only provide broad repair coverage but also contribute distinct, non-overlapping fixes.

This diversity suggests that ensemble or cooperative repair strategies leveraging multiple agentic models could improve overall coverage and robustness in multi-hunk repair.

\subsection{Multi-hunk Repair Complexity (RQ3)}
\label{sec:complexity}






 \begin{change}
  \begin{table*}
  \centering
  \scriptsize
  \caption{Hunk divergence for passing and failing repairs, and spatial proximity distribution (\% Pass) across coding agents on \benchmarkdataset }
  \label{tab:agent-performance-combined}
  \begin{tabular}{l|ccc|ccc|ccccc}
  \toprule
  \textbf{Agent}
    & \multicolumn{3}{c|}{\textbf{Hunk Divergence (Pass)}}
    & \multicolumn{3}{c|}{\textbf{Hunk Divergence (Fail)}}
    & \multicolumn{5}{c}{\textbf{Spatial Proximity (\% Pass)}} \\
  & Median & Mean & Max
    & Median & Mean & Max
    & Nucleus & Cluster & Orbit & Sprawl & Fragment \\
  \midrule
  \qwencode
    & 0.31 & 0.36 & 1.27
    & 0.45 & 0.49 & 1.60
    & 24.62 & 32.34 & 19.44 & 23.53 & 13.33 \\
  \geminicli
    & 0.36 & 0.38 & 1.27
    & 0.47 & 0.50 & 1.60
    & 49.23 & 49.75 & 26.39 & 35.29 & 33.33 \\
  \codex
    & 0.39 & 0.44 & 1.56
    & 0.47 & 0.53 & 1.60
    & 92.31 & 84.58 & 90.28 & 92.16 & 73.33 \\
  \claudecode
    & 0.40 & 0.44 & 1.56
    & 0.47 & 0.62 & 1.60
    & 100.00 & 93.03 & 87.50 & 92.16 & 86.67 \\
  \bottomrule
  \end{tabular}
  \end{table*}
  \end{change}
  
We next examine how repair accuracy varies with the structural complexity of multi-hunk bugs, measured through hunk divergence and spatial proximity (see Section \ref{subsec:hunk-characteristics}). Table~\ref{tab:agent-performance-combined} summarizes divergence statistics for passing and failing repairs, along with the proportion of successfully repaired cases in the five spatial proximity classes.


\begin{change}
Across all coding agents, passing repairs exhibit consistently lower hunk divergence than failing ones. For instance, the median divergence for fixed bugs ranges from 0.31 in \qwencode to 0.40 in \claudecode, whereas failing repairs show substantially higher values, with medians reaching 0.47 in \claudecode, \codex, and \geminicli. This pattern indicates that increasing lexical, structural, and file-level dispersion among hunks correlates with decreased repair accuracy. Agents struggle to coordinate edits when patches involve dissimilar or widely distributed code fragments, reinforcing the role of hunk divergence as a predictor of repair difficulty.
\end{change}


\begin{change}
The spatial proximity results further support this observation. Repair success tends to decline as bugs become more dispersed. \claudecode achieves 100\% repair accuracy in Nucleus bugs (single-method edits) and maintains high success rates across all categories: 93.03\% for Cluster (single-file), 87.50\% for Orbit (neighboring files), 92.16\% for Sprawl (distributed), and 86.67\% for Fragment (highly dispersed). \codex exhibits similar strength with 92.31\% (Nucleus), 84.58\% (Cluster), 90.28\% (Orbit), 92.16\% (Sprawl), and 73.33\% (Fragment). In contrast, \geminicli shows moderate accuracy ranging from 49.23\% (Nucleus) to 26.39\% (Orbit), while \qwencode achieves only 24.62\% (Nucleus), 32.34\% (Cluster), 19.44\% (Orbit), 23.53\% (Sprawl), and 13.33\% (Fragment). 

These results demonstrate that most repairs occur within localized categories, while highly dispersed Fragment bugs remain challenging, with only \claudecode and \codex exceeding 70\% accuracy on this category. Notably, \claudecode shows the strongest cross-file reasoning capability, maintaining consistently high repair accuracy even under dispersed editing requirements.
\end{change}

\begin{change}Overall, the results indicate that both divergence and spatial dispersion strongly influence the accuracy of agentic repair.\end{change}
As multi-hunk complexity increases, agents require deeper contextual understanding and more precise coordination across edit locations. Enhancing models with divergence-aware retrieval and proximity-sensitive context integration remains an important direction for improving the effectiveness of multi-hunk repair.



\begin{figure}[h]
  \centering
  \includegraphics[width=0.8\columnwidth]{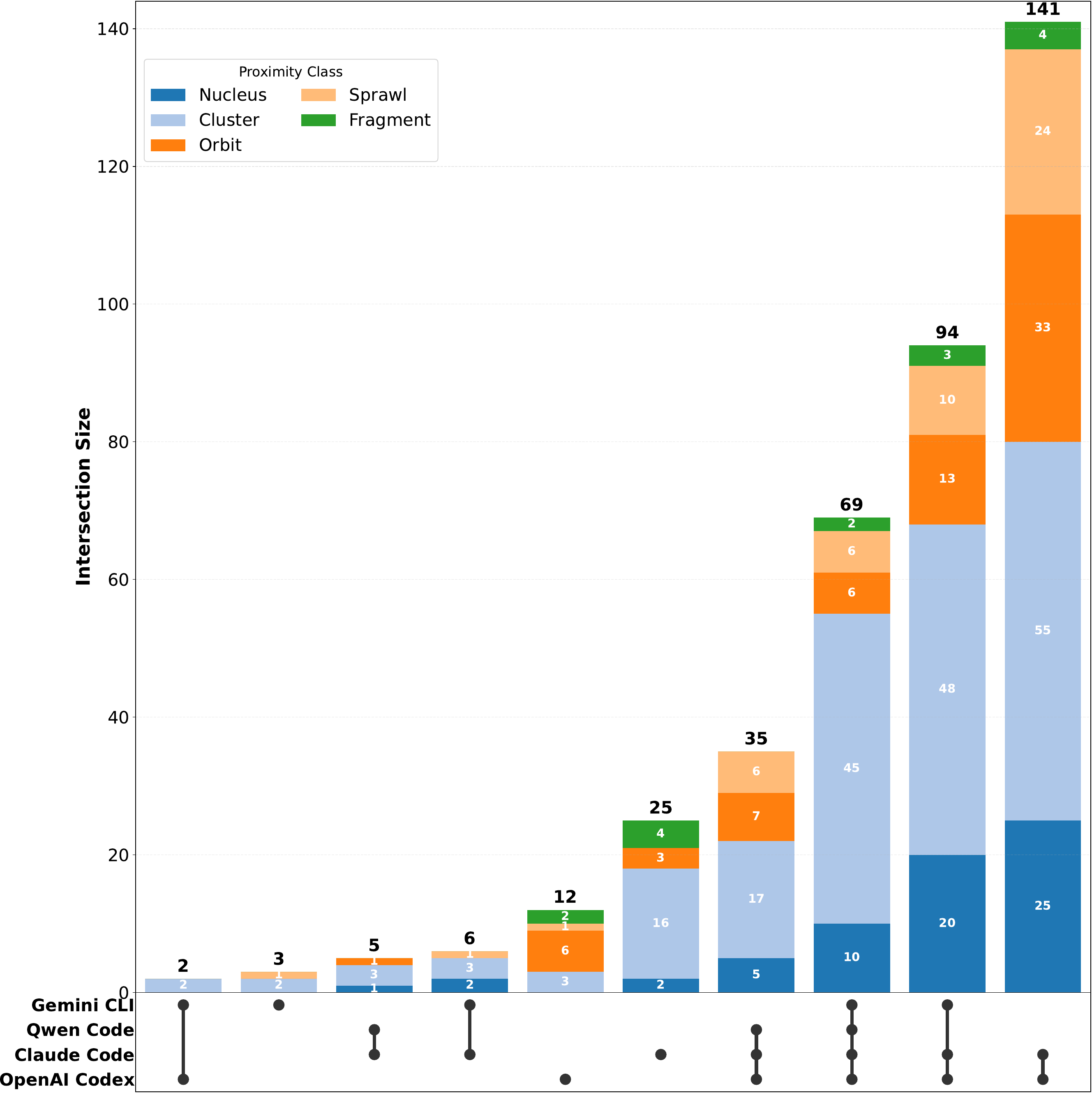}
  \caption{UpSet plot showing bugs correctly repaired by the selected coding agents, grouped by spatial proximity class. Bars indicate intersections among coding agents; color segments denote proximity categories.}  
  \label{fig:proximity-upset}
\end{figure}

\header{Intersections by Spatial Proximity} Figure~\ref{fig:proximity-upset} visualizes the intersection patterns of multi-hunk bugs repaired by the four coding agents, with each intersection decomposed by spatial proximity class.  Unlike the Venn diagram, this UpSet plot provides a more granular view of overlap structure, revealing how spatial dispersion interacts with agent collaboration. Each vertical bar represents a unique combination of agents, while the stacked color segments indicate the spatial distribution of fixed bugs—Nucleus (dark blue), Cluster (light blue), Orbit (orange), Sprawl (tan), and Fragment (green). The numbers above the bars denote the total number of bugs in each intersection.


\begin{change}
Smaller intersections, ranging from 2 to 12 bugs, correspond to isolated or agent-specific fixes, for instance, two bugs repaired jointly by \geminicli and \codex, and 12 repaired exclusively by \codex. Medium-sized overlaps capture more substantial collaboration: \claudecode alone repairs 25 bugs, and the combination of \qwencode, \claudecode, and \codex accounts for 35. The largest intersections reveal the most significant collaborative regions: 141 repairs shared by \claudecode and \codex alone, 94 bugs fixed by \geminicli, \claudecode, and \codex (excluding \qwencode), and 69 bugs repaired by all four agents.
\end{change}

\begin{change}
The spatial distribution within these intersections exposes a clear trend in repair difficulty.
Cluster-class bugs dominate across most groups, comprising 65\% of the 69 bugs fixed by all agents, followed by Nucleus (14\%), Orbit (9\%), Sprawl (9\%), and Fragment (3\%).
Fragment-class bugs remain rare, appearing mainly in the \claudecode-only intersection and the \claudecode and \codex pair.
Bugs with higher proximity, those in the Cluster and Nucleus categories, are consistently solvable across models, while spatially dispersed bugs (Orbit, Sprawl, and Fragment) are handled by fewer agents and require stronger reasoning capabilities.
\end{change}


\begin{change}
Overall, Cluster-class bugs dominate across intersections, indicating that spatially localized repairs are consistently solvable by most agents. In contrast, Fragment-class bugs are relatively rare, appearing only in limited intersections, such as those unique to \claudecode and the overlap between \claudecode and \codex, underscoring the difficulty of highly dispersed repair scenarios. The largest consensus group, comprising 69 bugs repaired by all four agents, represents the most stable subset of multi-hunk bugs and reflects tasks consistently solved by current coding agents. Additionally, the three-agent overlap excluding \qwencode, which accounts for 94 bugs, demonstrates that \claudecode, \codex, and \geminicli collectively handle a broader range of complex, multi-hunk scenarios. Finally, Nucleus-class bugs appear consistently across intersections, suggesting that moderate spatial dispersion remains within the capability of most agents.
\end{change}


\begin{figure}
    \centering
    \includegraphics[width=0.8\columnwidth]{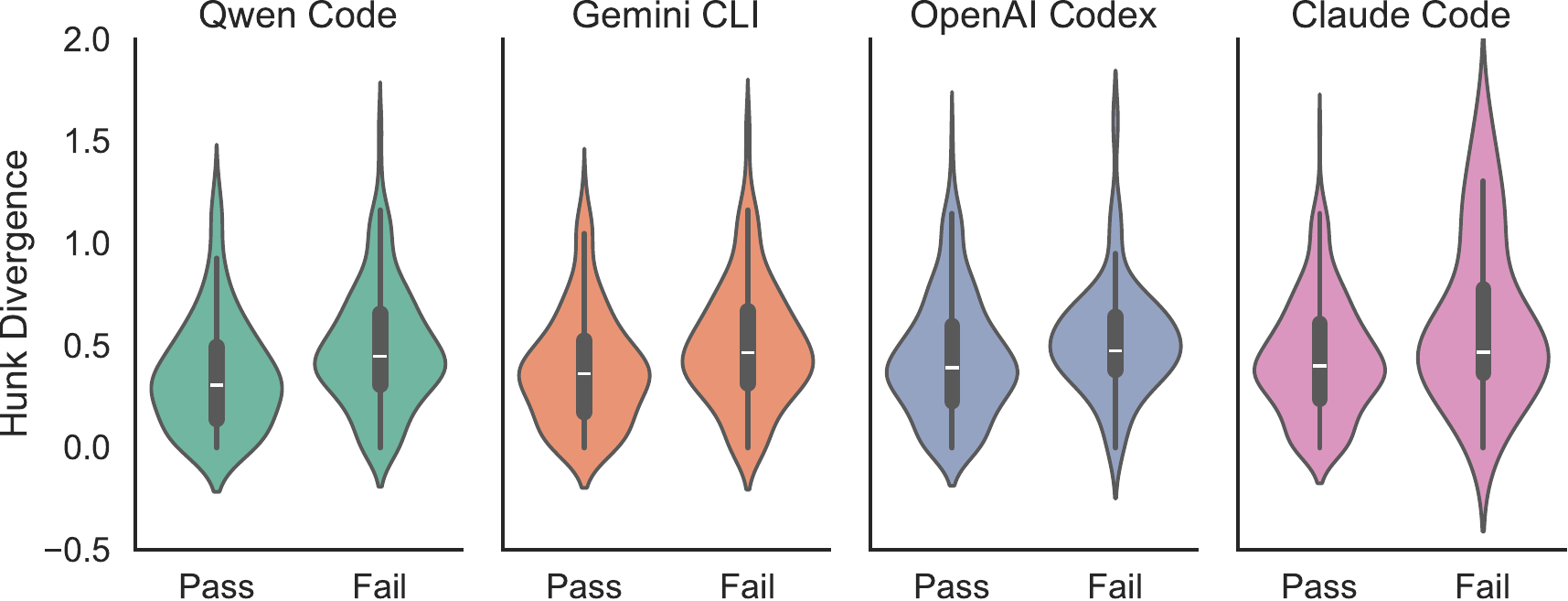}
    \caption{Faceted violin plots of hunk divergence values, grouped by coding agents and split by outcome.}    
    \label{fig:llm_divergence_violin}
\end{figure}

\header{Hunk Divergence as a Predictor of Agentic Repair Success} Figure~\ref{fig:llm_divergence_violin} presents a faceted violin plot showing the distribution of hunk divergence values for bugs that were successfully repaired (Pass) versus those that were not (Fail) across the four coding agents. \changeinline{Each facet corresponds to an agent, \qwencode, \geminicli, \codex, and \claudecode, with the y-axis representing the divergence of the hunk on the same scale across facets.} Within each facet, the embedded boxplots show the medians and quartile ranges.

\begin{change}
Across all agents, a consistent trend is observed: successful repairs are concentrated at lower divergence values, with median divergence for passing cases ranging from 0.31 to 0.40. In contrast, failed repair attempts exhibit higher divergence, with the median divergence of failing cases ranging from 0.45 to 0.47 and a long tail extending up to 1.6.
For \qwencode, passing bugs have a median divergence of 0.31, with an interquartile range of 0.14 to 0.49, whereas failed repairs have a median of 0.45 and can reach 1.6.
\geminicli follows a similar pattern, with pass cases centered at 0.36 (extending to 1.27) and fail cases at 0.47 (extending to 1.6).
\codex shows a slightly higher pass median of 0.39 and a fail median of 0.47, though its fail distribution is narrower, aligning with its higher overall repair accuracy. 
\claudecode shows the largest separation between passing and failing cases in terms of mean divergence (0.44 for passing vs. 0.62 for failing). For passing cases, the median divergence is 0.40 with an interquartile range of 0.24–0.61. In contrast, the 29 failing cases have a higher median divergence of 0.47, a wider interquartile range of 0.41, and a long upper tail extending to 1.6.
This wide distribution of failing cases, with values extending to high divergence, indicates that \claudecode successfully repairs most low-divergence bugs but is less effective on bugs that require highly fragmented patches.
\end{change}

Overall, the results show a consistent relationship between repair success and hunk divergence. Bugs with lower divergence values, corresponding to more compact and semantically aligned edits, are more likely to be successfully repaired across agents. Bugs requiring more dispersed or heterogeneous changes are less likely to be repaired. These results indicate that hunk divergence is a strong predictor of repair success in agentic tools, underscoring that spatial and structural coherence across edits remains critical for effective multi-hunk repair.

 \begin{change}
  \begin{table*}[t]
  \centering
  \scriptsize
  \caption{Characteristics of bugs not fixed by all four agents on \benchmarkdataset}
  \label{tab:unsolved-bugs-comprehensive}
  {
  \begin{tabular*}{\textwidth}{@{\extracolsep{\fill}}lccccrrrl@{}}
  \toprule
  \textbf{Bug ID} & \multicolumn{4}{c}{\textbf{Localization Success}} & \multicolumn{3}{c}{\textbf{Multi-Hunk Complexity}} & \\
  \cmidrule(lr){2-5} \cmidrule(lr){6-8}
   & \textbf{\qwencode} & \textbf{\geminicli} & \textbf{\codex} & \textbf{\claudecode} & \textbf{Hunks} & \textbf{Divergence} & \textbf{Proximity} & \\
  \midrule
  \code{JacksonDatabind\_104} & \cmark & \xmark & \cmark & \xmark & 3  & 0.463 & Cluster & \\
  \code{JacksonDatabind\_105} & \xmark & \xmark & \xmark & \xmark & 2  & 0.588 & Cluster & \\
  \code{JacksonDatabind\_108} & \xmark & \xmark & \cmark & \xmark & 2  & 0.268 & Cluster & \\
  \code{JacksonDatabind\_110} & \cmark & \cmark & \cmark & \xmark & 3  & 0.584 & Cluster & \\
  \code{JacksonDatabind\_80}  & \xmark & \xmark & \cmark & \xmark & 3  & 0.414 & Cluster & \\
  \code{JacksonDatabind\_81}  & \xmark & \xmark & \cmark & \xmark & 7  & 1.158 & Cluster & \\
  \code{Astropy-7606}         & \cmark & \cmark & \cmark & \cmark & 2  & 0.141 & Cluster & \\
  \code{Closure\_171}         & \xmark & \xmark & \xmark & \xmark & 2  & 0.363 & Orbit   & \\
  \code{JacksonDatabind\_109} & \xmark & \xmark & \xmark & \xmark & 5  & 0.702 & Orbit   & \\
  \code{Astropy-8707}         & \xmark & \cmark & \cmark & \cmark & 6  & 0.450 & Orbit   & \\
  \code{Closure\_157}         & \xmark & \xmark & \xmark & \xmark & 7  & 1.078 & Sprawl  & \\
  \code{JacksonDatabind\_103} & \xmark & \xmark & \xmark & \xmark & 26 & 1.599 & Sprawl  & \\
  \midrule
   & \multicolumn{4}{c}{\textbf{Rate}} & \multicolumn{2}{c}{\textbf{Mean}} & & \\
  \cmidrule(lr){2-5} \cmidrule(lr){6-7}
  \addlinespace[2pt]
  \textbf{12 Non-fixed Bugs} & 0.25  & 0.25  & 0.58  & 0.17   & 5.67 & 0.651 & --- & \\
  \textbf{All 404 Bugs}      & 0.40  & 0.49  & 0.75  & 0.66   & 3.84 & 0.452 & --- & \\
  \addlinespace[2pt]
  \textbf{Diff ($\Delta$)}   & -38\% & -49\% & -22\% & -75\%  & +48\% & +44\% & --- & \\
  \bottomrule
  \end{tabular*}
  }
  \end{table*}
  \end{change}

\header{Failure Case Analysis: Bugs Unfixed by Any Agent} While the majority of bugs were successfully repaired by at least one agent, \changeinline{12} bugs (\changeinline{2.97\%} of the dataset) remained unsolved by all four agents. We conducted a comprehensive analysis of these failure cases to understand the fundamental limitations of current agentic repair approaches. Our analysis reveals several findings that challenge conventional assumptions in automated program repair with coding agents. The analysis for these bugs is shown in table~\ref{tab:unsolved-bugs-comprehensive}.

We first analyze localization accuracy by examining whether agents correctly identified the files requiring modification. Surprisingly, \changeinline{58.33\% of unsolved bugs (7 out of 12)} were successfully localized by at least one agent. Specifically, \codex successfully localized \changeinline{7} bugs, \qwencode and \geminicli each localized \changeinline{3} bugs, and \claudecode localized \changeinline{2}. Notably, bug \code{JacksonDatabind\_110} was correctly localized by three different agents (\geminicli, \qwencode, and \codex), yet none could generate a correct repair. This reveals a distinct \emph{localization-repair gap} that agents can identify where the bug is located, but fail to synthesize an appropriate fix.


\begin{change}
We also observe that unsolved bugs exhibit significantly higher complexity as measured by hunk divergence. The mean hunk count for unsolved bugs is 5.67 compared to 3.84 for the overall dataset, a 48\% increase. Also, a surprising finding concerns the spatial distribution of changes. We classified bugs by proximity class based on how close the required changes are to each other. Intuitively, bugs requiring spatially close changes (classified as ``Cluster'') should be easier to fix because related modifications are localized. However, 58.33\% of unsolved bugs are Cluster-type, with a mean divergence of 0.52, higher than the overall mean of 0.452. This analysis reveals that physical co-location of changes does not guarantee a successful repair by coding agents. Despite changes occurring in the same file or nearby lines, high divergence indicates that agents struggle to generate a coherent fix.
\end{change}


\begin{change}
The distribution of unsolved bugs across projects is highly skewed. Eight of the twelve unsolved bugs (66.67\%) originate from the \code{JacksonDatabind} project, despite \code{JacksonDatabind} bugs comprising only 10.15\% of the overall dataset. The remaining four unsolved bugs are split between \code{Closure} (two bugs, Java) and \code{Astropy} (two bugs, Python). \code{JacksonDatabind} bugs are particularly challenging because they involve complex type systems with deep inheritance hierarchies, annotation-driven behavior where semantics are specified through metadata rather than code, and polymorphic serialization logic requiring runtime type resolution.
\end{change}

Overall, despite localization success for \changeinline{7 of the 12} unsolved bugs, none were successfully repaired, demonstrating a distinct localization-repair gap. The high mean hunk count and divergence values reveal a complexity threshold beyond which all agents fail.


\begin{figure}[h]
    \centering
    \includegraphics[width=0.7\columnwidth]{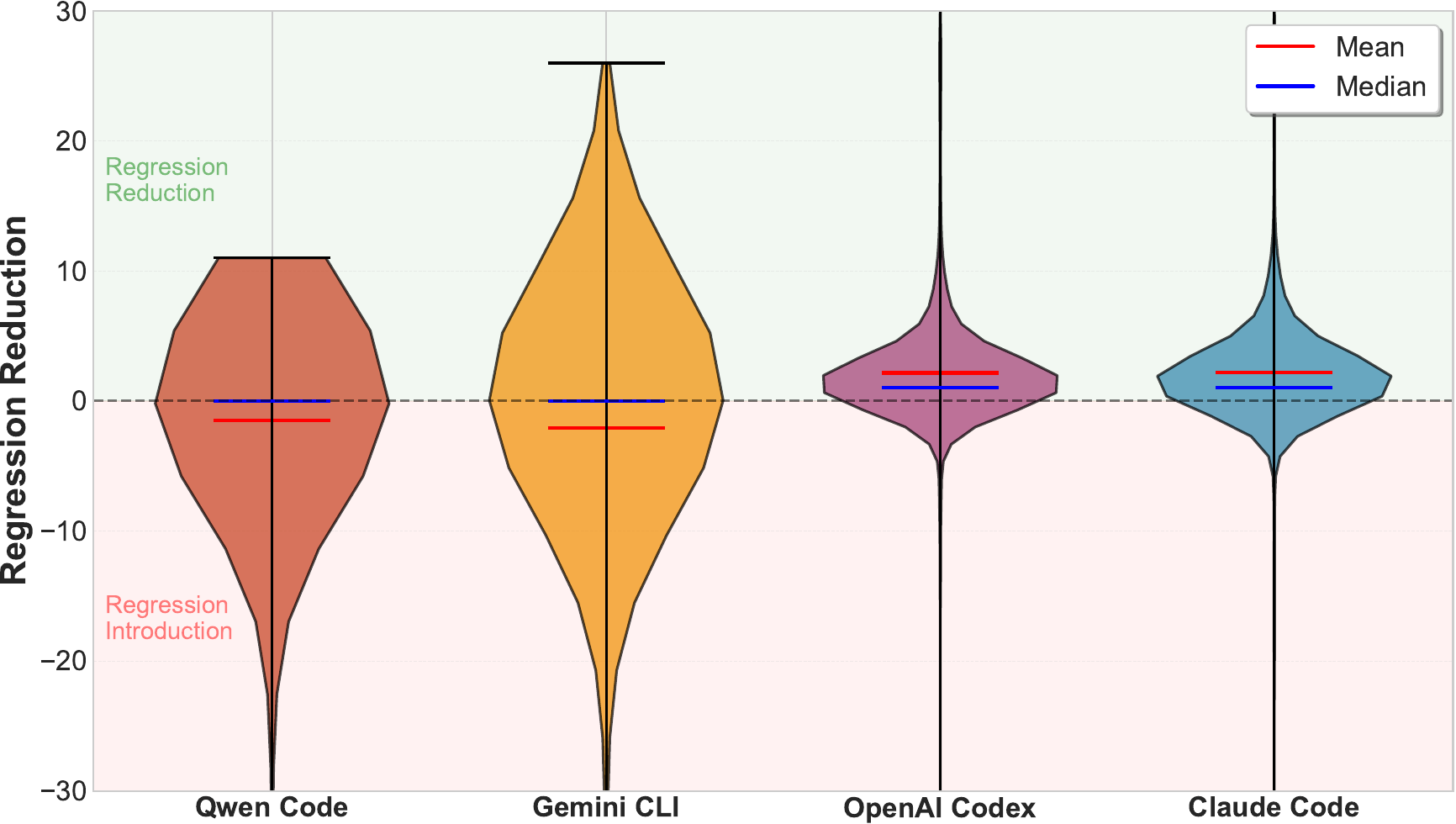}
    \caption{Regression reduction across coding agents}
    \label{fig:agent-regression-reduction-violin}
\end{figure}

\subsection{Regression (RQ4)}


\begin{change}
Table~\ref{tab:multiagent-accuracy-combined} presents regression reduction statistics across all four coding agents. Our results reveal a stark divide in effectiveness. \claudecode and \codex both achieve a positive regression reduction (+2.17 and +2.16, respectively). In contrast, \qwencode and \geminicli exhibit negative regression reduction (-1.50 and -2.10, respectively), introducing an average of 1--2 new test failures per repair attempt.
\end{change}


\begin{change}
\claudecode and \codex achieve both the highest repair accuracy and the most substantial reduction in regressions.
\qwencode and \geminicli exhibit the lowest repair accuracy at 26.98\% and 43.07\%. The same two agents also exhibit the most negative regression reduction values at -1.50 and -2.10. \end{change} Although the sample size is small, our observation suggests that higher repair rates do not sacrifice code quality for quantity. This suggests that effective bug repair requires not aggressive code generation, but rather \textit{context-aware} and \textit{dependency-conscious} program synthesis. Tools exhibiting positive regression reduction demonstrate a deeper understanding of code structure and test relationships.


In real-world deployment, the introduction of regression is particularly problematic. A ``repair'' that fixes one bug while introducing new failures creates net negative value, imposing a \textit{regression tax} on development teams who must debug both the original issue and the newly introduced failures. Our results suggest that \qwencode and \geminicli, despite achieving non-trivial repair rates (\changeinline{26.98\%} and \changeinline{43.07\%}), would require extensive human review before deployment in production. The repair success rate alone provides an incomplete and potentially misleading evaluation of tool quality. For example, Tool~A achieves a 60\% repair success rate with an average regression change of -2.0, while Tool~B achieves a 50\% repair success rate with an average regression improvement of +1.0. This example shows that evaluations should be more nuanced and take into account the overall practical value of using agentic coding tools.

\header{Variability in Regression Reduction} Figure~\ref{fig:agent-regression-reduction-violin} presents violin plots that reveal the complete probability density of regression reduction outcomes. The width at any vertical position represents probability density; wider sections indicate common outcomes, narrower sections indicate rare outcomes. Red and blue lines mark the mean and median, respectively.

The violin shapes reveal a stark categorical divide across coding agents. \claudecode achieves \changeinline{92.6}\% positive outcomes (\changeinline{374} of \changeinline{404} repairs), \changeinline{5.7}\% zero-change (\changeinline{23} repairs), and only \changeinline{1.7}\% negative (\changeinline{7} repairs). \codex shows similar behavior with \changeinline{90.0}\% positive, \changeinline{7.5}\% zero-change, and \changeinline{2.5}\% negative. Both exhibit narrow distributions with low variance ($\sigma=\changeinline{7.02}$ and $\changeinline{6.39}$), indicating predictable behavior. Their worst-case scenarios are bounded (min $\changeinline{-80}$ and $-47$). In contrast, \qwencode achieves only \changeinline{31.6}\% positive outcomes (\changeinline{125} of \changeinline{396} repairs), with \changeinline{58.1}\% zero-change (\changeinline{230} repairs compile but produce no improvement) and \changeinline{10.4}\% negative (\changeinline{41} repairs introduce regressions). \geminicli shows \changeinline{47.8}\% positive, \changeinline{40.9}\% zero, and \changeinline{11.3}\% negative. Both display wide, dispersed distributions with high variance ($\sigma=\changeinline{27.71}$ and $\changeinline{30.25}$), \changeinline{around 4}$\times$ larger than Claude/Codex—and catastrophic worst-cases (min $-543$ and $-488$), where a single repair breaks hundreds of tests.

The variance ratio ($\sigma_{\text{Gemini}}^2 / \sigma_{\text{Claude}}^2 \approx \changeinline{19}$) quantifies operational risk: high-variance tools require substantially more quality assurance resources per repair. Distribution shape predicts automation feasibility—narrow violins enable autonomous deployment in CI/CD pipelines, while wide violins necessitate supervised deployment with mandatory human review. \geminicli's range $[-488, +26]$ includes catastrophic outcomes that could halt development, whereas \claudecode's $[\changeinline{-80}, +73]$ contains only manageable scenarios. This tail risk asymmetry explains why practitioners need to be careful while leveraging high-variance coding assistance tools despite seemingly acceptable averages. When selecting tools to build production-ready applications, practitioners must consider different aspects: central tendency (mean, median) indicates typical outcomes, dispersion (standard deviation) reveals consistency, and worst-case tail risk (minimum/maximum values) exposes catastrophic failures.

Overall, \emph{regression reduction} reveals the hidden cost of automated repairs. Tools with negative regression reduction impose a \emph{regression tax} on development teams, potentially creating more problems than they solve. This aspect requires careful examination when evaluating future coding agents for software engineering tasks.

\subsection{Operational Dynamics (RQ5)}

\subsubsection{Token Consumption}


  \begin{change}
  \begin{table*}[t]
  \centering
  \scriptsize
  \caption{Token consumption per bug on multi-hunk bugs with coding agents}
  \label{tab:token-usage-statistics-combined}
  \begin{tabular}{lrrrr}
  \toprule
  & \multicolumn{2}{c}{\textbf{Pass}} & \multicolumn{2}{c}{\textbf{Fail}} \\
  \cmidrule(lr){2-3} \cmidrule(lr){4-5}
  \textbf{Agent} & \textbf{Input} & \textbf{Output} & \textbf{Input} & \textbf{Output} \\
  \midrule
  \qwencode    & 1.61M  & 6.93K  & 2.32M   & 9.56K  \\
  \geminicli   & 1.23M  & 3.85K  & 6.66M   & 12.82K \\
  \codex       & 91.80K & 5.77K  & 121.60K & 14.77K \\
  \claudecode  & 799    & 17.51K & 245     & 27.15K \\
  \bottomrule
  \end{tabular}
  \end{table*}
  \end{change}

While repair success measures functional correctness, it overlooks a critical operational dimension: token consumption. We analyze token consumption patterns across agents to understand the computational cost of repair attempts and whether successful repairs differ from failed ones in token consumption.

We measure API token consumption in two dimensions. Input tokens represent the total tokens sent to the model across all repair attempts. Output tokens represent the total tokens generated by the model across all repair attempts. For \qwencode, \geminicli, and \claudecode, token counts are extracted directly from agent trajectory logs and aggregated. For \codex, which does not log token counts, we parse all input prompts and output responses using tiktoken~\footnote{\url{https://github.com/openai/tiktoken}} to compute token count, then aggregate across all repair attempts.

\begin{figure}[t]
\centering
\includegraphics[width=0.8\columnwidth]{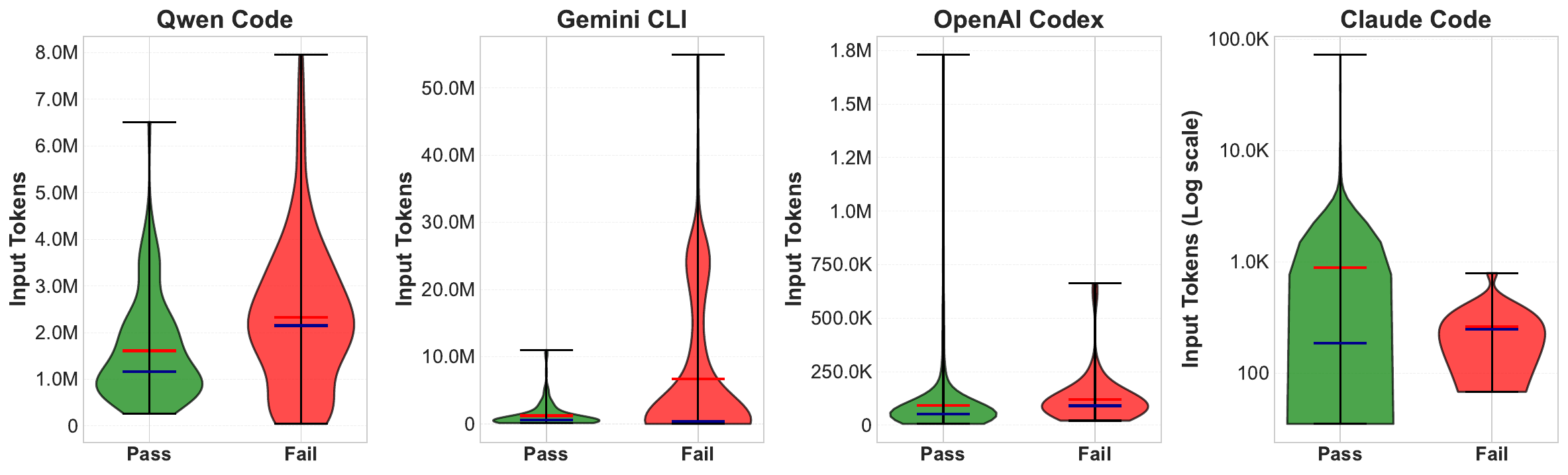}
\caption{Violin plot distributions of input token consumption across coding agents. Each agent has independent y-axis scaling.}
\label{fig:token-distribution-input}
\end{figure}
\begin{figure}[t]
\centering
\includegraphics[width=0.8\columnwidth]{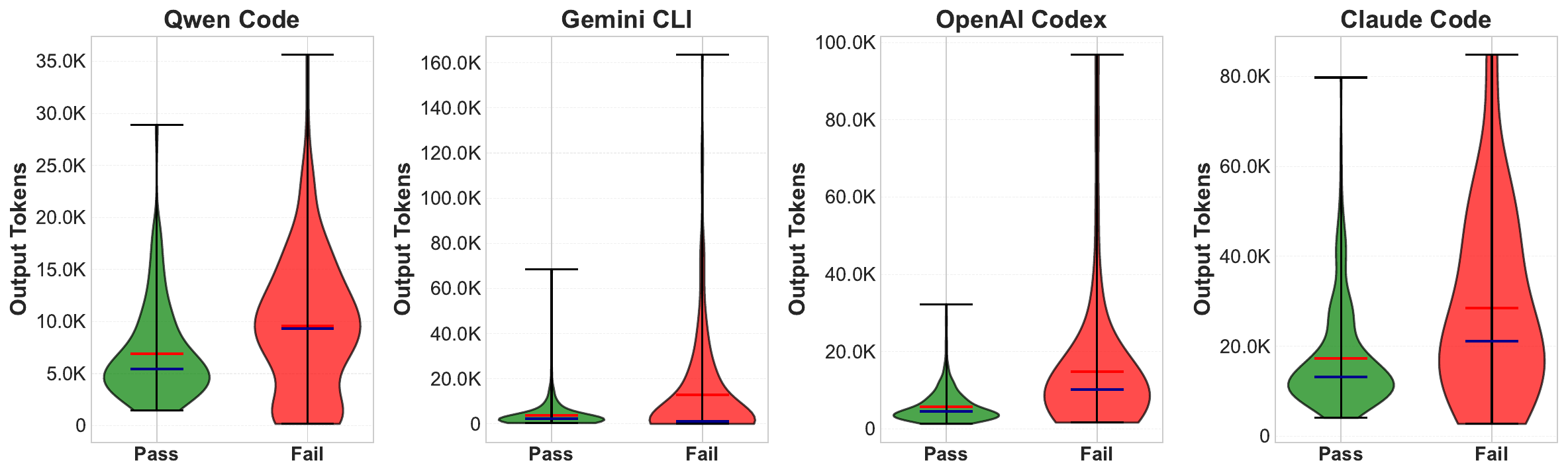}
\caption{Violin plot distributions of API output token generation across coding agents. Each agent has independent y-axis scaling.}
\label{fig:token-distribution-output}
\end{figure}

\header{Token Consumption Across Agents} Table~\ref{tab:token-usage-statistics-combined} presents token consumption statistics across all four agents and Figures~\ref{fig:token-distribution-input}--\ref{fig:token-distribution-output} show violin plots of token distributions by outcome. \begin{change} \claudecode demonstrates the lowest per-bug input token consumption, representing 115$\times$ less than \codex (91.80K), 1,540$\times$ less than \geminicli (1.23M), and 2,015$\times$ less than \qwencode (1.61M). The prompt caching mechanism in \claudecode enables this reduction by reusing previously cached project context. For output tokens per bug, \geminicli generates the fewest at 3.85K tokens on average, followed by \codex at 5.77K, \qwencode at 6.93K, and \claudecode at 17.51K.
\end{change}



\begin{figure}[t]
\centering
\includegraphics[width=0.8\columnwidth]{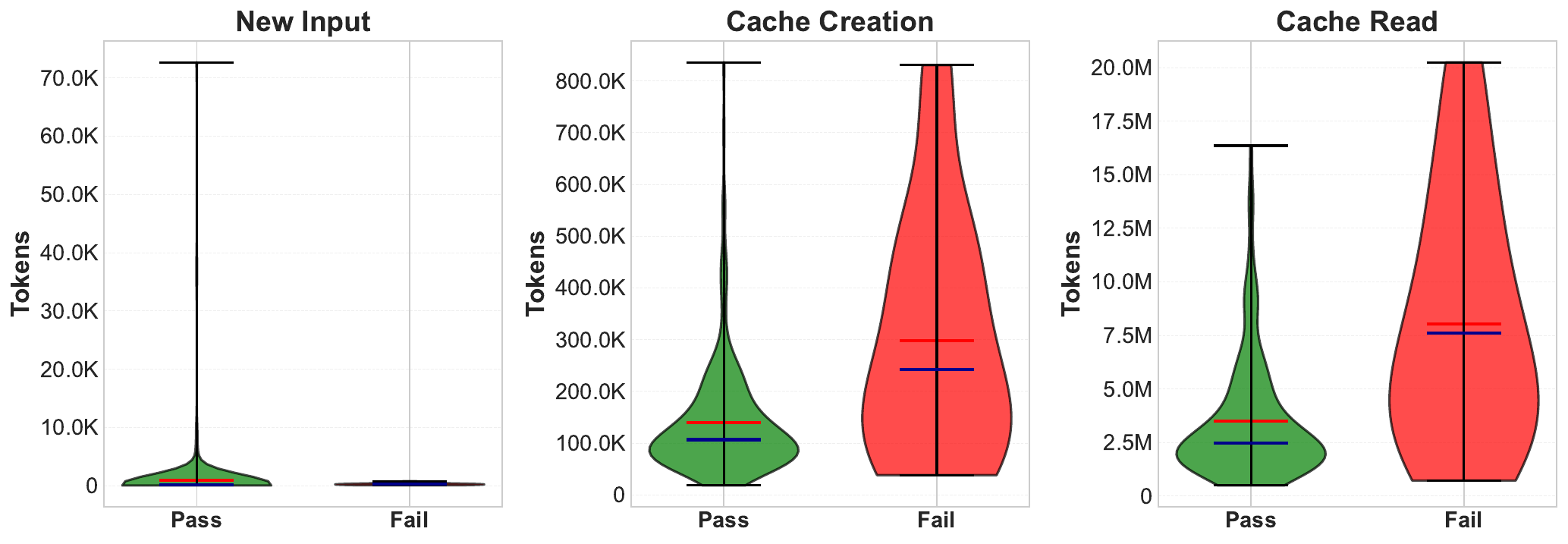}
\caption{\claudecode input token distribution by type and repair outcome.}
\label{fig:claude-input-token-distribution-violin}
\end{figure}

Figure~\ref{fig:claude-input-token-distribution-violin} shows new input tokens only for \claudecode (not total input including cache). \claudecode's prompt caching divides input tokens into three distinct categories. New input tokens represent uncached prompt content unique to each request, such as bug-specific information, test failures, and iterative repair instructions (successful repairs average \changeinline{799} per bug). Cache creation tokens represent prompt content written to the cache for reuse across multiple API calls, such as project source code, dependency files, and static context (successful repairs average \changeinline{141K} per bug). Cache read tokens represent previously cached content retrieved in subsequent API invocation (successful repairs average \changeinline{3.62M} per bug). Figure~\ref{fig:claude-input-token-distribution-violin} reveals that cache read tokens dominate at 96\% of total input volume. 

\header{Token Consumption by Repair Outcome} Failed repairs exhibit distinct token consumption patterns across agents, as shown by wider failure distributions (red violins) in Figure~\ref{fig:token-distribution-input} and Figure~\ref{fig:token-distribution-output}. \qwencode's failed repairs use \changeinline{44.1\%} more input tokens and \changeinline{38.0\%} more output tokens compared to successful repairs. \geminicli exhibits the most dramatic increase, with failed repairs consuming \changeinline{440.4\%} more input tokens and \changeinline{232.8\%} more output tokens. \codex shows \changeinline{32.5\%} more input tokens and \changeinline{156.0\%} more output tokens for failures. These patterns indicate that failures do not terminate early; agents persist through multiple unsuccessful attempts before giving up.

\claudecode shows a distinct pattern due to its prompt caching mechanism. Failed repairs use \changeinline{69.4\%} fewer new input tokens (\changeinline{245 vs.\ 799}) but generate \changeinline{55.1\%} more output tokens (\changeinline{27.2K vs.\ 17.5K}). The lower number of new input tokens results from cache reuse across repeated repair attempts, where most of the prompt context remains unchanged. Figure~\ref{fig:claude-input-token-distribution-violin} shows that failed repairs consume \changeinline{87.6\%} more cache creation tokens and \changeinline{110.5\%} more cache read tokens than successful ones. This indicates that caching enables the reuse of stored context and avoids the cost of resending large inputs. Overall, prompt caching reduces expensive input token usage while preserving access to the required contextual information.

\begin{change}
    \begin{table*}[t]
    \scriptsize
    \centering
    \scriptsize
    \caption{Cost per bug on multi-hunk bugs with coding agents}
    \label{tab:cost-effectiveness}
    \small
    \begin{tabular}{lrrrrr}
    \toprule
    \textbf{Agent} & \textbf{Pass cost} & \textbf{Fail cost} & \textbf{Fail/Pass} & \textbf{Total cost} & \textbf{Cost / repair} \\
                   & (\$/bug)           & (\$/bug)           & (ratio)            & (\$, 404 bugs)      & (\$)                   \\
    \midrule
    \qwencode   & 0.49 & 0.62 & 1.26$\times$ & 236.52 & 2.17 \\
    \geminicli  & 0.37 & 2.02 & 5.42$\times$ & 529.95 & 3.05 \\
    \codex      & 0.17 & 0.30 & 1.74$\times$ &  76.16 & 0.22 \\
    \claudecode & 1.88 & 3.69 & 1.96$\times$ & 812.06 & 2.17 \\
    \bottomrule
    \end{tabular}
    \end{table*}

\header{Cost Analysis}
We estimate monetary cost using November 2025 pricing\footnote{\textsc{Claude Sonnet~4.5}: \$3/\$15 per MTok input/output, cache write \$3.75/MTok, cache read \$0.30/MTok. \textsc{GPT-5}: \$1.25/\$10. \textsc{Gemini-2.5-Flash}: \$0.30/\$2.50. \textsc{qwen3-coder-flash} via OpenRouter: \$0.30/\$1.20.}, combined with per-bug token averages (Table~\ref{tab:token-usage-statistics-combined}) and pass/fail counts (Table~\ref{tab:multiagent-accuracy-combined}). 

\codex produces a successful repair at an average cost of \$0.22. This is approximately $10\times$ lower than \claudecode and \qwencode (both \$2.17), and about $14\times$ lower than \geminicli\ (\$3.05), primarily because \codex uses substantially fewer input tokens. Failed attempts cost between 1.3$\times$ and 5.4$\times$ more than successful ones (Table~\ref{tab:cost-effectiveness}), and also take longer (Table~\ref{tab:runtime-success-failure-combined}). This cost disparity is most pronounced for \geminicli, where failed repairs account for 87.7\% of the total cost (\$529.95) and have longer runtimes than successful repairs (471.53\,s vs.\ 151.89\,s).

\claudecode achieves the highest accuracy but also the highest total cost (\$812.06), while \codex has the lowest cost per successful repair.
\end{change}

\subsubsection{Runtime} 


  \begin{change}
  \begin{table*}[t]
  \centering
  \scriptsize
  \caption{Mean runtime (in seconds) for successful and failed repairs across coding agents}
  \label{tab:runtime-success-failure-combined}
  \begin{tabular}{lrrr}
  \toprule
  \textbf{Agent} & \textbf{Successful Repairs (s)} & \textbf{Failed Repairs (s)} & \textbf{Pass-Fail Gap (s)} \\
  \midrule
  \qwencode        & 334.79 & 450.79  & 116.00  \\
  \geminicli       & 151.89 & 471.53  & 319.64  \\
  \codex           & 309.59 & 1331.69 & 1022.10 \\
  \claudecode      & 346.02 & 641.61  & 295.59  \\
  \midrule
  \textbf{Average} & 285.57 & 723.91  & 438.33  \\
  \bottomrule
  \end{tabular}
  \end{table*}
  \end{change}
  

\begin{change}
Table~\ref{tab:runtime-success-failure-combined} summarizes the mean runtime of successful and failed repair attempts across coding agents. 
Overall, failed repairs consistently require substantially more time than successful ones, with an average runtime of 723.91~s compared to 285.57~s. \geminicli exhibits the shortest runtime for successful repairs (151.89~s), while \codex exhibits the longest runtime for failed repairs (1331.69~s). \qwencode demonstrates the smallest pass--fail gap (116.00~s), indicating relatively stable runtime behavior across outcomes.
\end{change}


\begin{change}
Failed repairs consistently require substantially longer runtimes than successful ones across all agents. This pattern suggests that agents continue working on failed attempts before terminating. Agents differ in their level of persistence. \codex continues for the longest, with an average of 1331.69~s before terminating a failed attempt. \qwencode terminates earliest, with an average of 450.79~s.
\end{change}

\subsubsection{Understanding agentic actions}

The actions $a \in \mathcal{A}$ (Equation~\ref{eq:tool-calls}) can be implemented in two distinct ways. First, coding agents may invoke \emph{native tools} provided directly by the agent framework. For example, \claudecode provides \code{read\_file} to retrieve file content, \code{edit} to modify code, and \code{search\_file\_content} to search within files. Similarly, \qwencode offers \code{write\_file} to edit files, \code{list\_directory} to enumerate directory contents, and \code{search\_file\_content} to locate code patterns. Second, agents may execute \emph{shell-invoked commands} through bash interfaces, including standard Unix utilities (\code{cat}, \code{sed}, \code{grep}), build systems (\code{mvn test}, and domain-specific frameworks (\code{defects4j compile}, \code{defects4j test}).


  \begin{change}
  \begin{table*}[t]
  \centering
  \scriptsize
  \caption{Unique command diversity across coding agents}
  \label{tab:unique_commands_per_agent-combined}
  \begin{tabular}{lrrr}
  \toprule
  \textbf{Agent} & \textbf{Native Tools} & \textbf{Shell-Invoked} & \textbf{Total Unique} \\
  \midrule
  \qwencode    & 14 & 85  & 99  \\
  \geminicli   & 11 & 32  & 43  \\
  \codex       & 0  & 102 & 102 \\
  \claudecode  & 7  & 67  & 74  \\
  \bottomrule
  \end{tabular}
  \end{table*}
  \end{change}
  
\header{Tool Diversity}
Table~\ref{tab:unique_commands_per_agent-combined} presents the number of unique tools invoked by each agent across all repair attempts in \benchmarkdataset. The total unique count represents distinct commands used at least once during evaluation.

Notably, \codex relies exclusively on shell-invoked commands (\changeinline{102} unique commands), executing every action through shell interfaces without native tool support. This terminal-centric design reflects a general-purpose interaction model where file operations, code editing, build execution, and version control are all performed via command-line invocations. \qwencode and \claudecode exhibit similar patterns, using primarily shell-invoked commands (\changeinline{85} and \changeinline{67} unique commands, respectively) with minimal native tool integration (\changeinline{14} and 7 native tools). In contrast, \geminicli demonstrates more balanced usage, employing 11 native tools alongside \changeinline{32} shell-invoked commands.

These differences in tool diversity and implementation paradigms have implications for how agents allocate effort across repair activities. Agents relying on shell-invoked commands must construct command-line invocations and interpret unstructured output, whereas agents leveraging native tools benefit from structured interfaces and reduced parsing overhead.

  \begin{change}
  \begin{table*}[t]
  \centering
  \scriptsize
  \caption{Tool category usage: successful and failed repairs across coding agents}
  \label{tab:all_agents_tool_category_pass_fail-combined}
  \setlength{\tabcolsep}{4pt}
  \begin{tabular}{l|rrr|rrr|rrr|rrr}
  \toprule
  \multirow{2}{*}{\textbf{Category}}
    & \multicolumn{3}{c|}{\textbf{\qwencode}}
    & \multicolumn{3}{c|}{\textbf{\geminicli}}
    & \multicolumn{3}{c|}{\textbf{\codex}}
    & \multicolumn{3}{c}{\textbf{\claudecode}} \\
  \cline{2-13}
   & \textbf{\makecell{Pass \\ (\%)}} & \textbf{\makecell{Fail \\ (\%)}} & \textbf{\makecell{Total \\ (\%)}}
   & \textbf{\makecell{Pass \\ (\%)}} & \textbf{\makecell{Fail \\ (\%)}} & \textbf{\makecell{Total \\ (\%)}}
   & \textbf{\makecell{Pass \\ (\%)}} & \textbf{\makecell{Fail \\ (\%)}} & \textbf{\makecell{Total \\ (\%)}}
   & \textbf{\makecell{Pass \\ (\%)}} & \textbf{\makecell{Fail \\ (\%)}} & \textbf{\makecell{Total \\ (\%)}} \\
  \midrule
  WRITE          & 15.1 & 15.7 & 15.6 & 25.8 & 42.9 & 38.0 & 44.8 & 41.6 & 44.1 & 22.6 & 17.6 & 22.0 \\
  READ           & 25.4 & 23.0 & 23.6 & 26.6 & 21.6 & 23.1 & 5.4  & 4.9  & 5.3  & 22.9 & 23.8 & 23.0 \\
  TEST           & 16.3 & 15.2 & 15.5 & 23.2 & 17.5 & 19.2 & 15.3 & 14.9 & 15.2 & 14.9 & 15.1 & 14.9 \\
  BUILD          & 8.4  & 10.1 & 9.7  & 12.3 & 10.9 & 11.3 & 6.0  & 10.0 & 6.8  & 11.7 & 11.6 & 11.7 \\
  SEARCH\_CONTENT & 9.1  & 8.1  & 8.4  & 2.1  & 1.7  & 1.8  & 9.4  & 10.3 & 9.6  & 8.9  & 14.6 & 9.5  \\
  SEARCH\_FILES   & 6.5  & 6.3  & 6.4  & 4.5  & 1.9  & 2.6  & 5.9  & 4.2  & 5.5  & 10.5 & 7.2  & 10.1 \\
  NAVIGATE       & 15.4 & 17.0 & 16.5 & 0.0  & 0.0  & 0.0  & 0.0  & 0.0  & 0.0  & 0.8  & 1.2  & 0.9  \\
  \bottomrule
  \end{tabular}
  \end{table*}
  \end{change}

\header{Tool Usage Patterns in Successful vs Failed Repairs} To analyze how agents allocate effort differently in successful versus unsuccessful repairs, we abstract all tool invocations within the action space $\mathcal{A}$ (Equation~\ref{eq:tool-calls}) into functional categories based on operational purpose. Commands are grouped by their function rather than implementation: \code{read\_file} and \code{cat} both constitute READ operations, while \code{edit} and \code{write\_file} both represent WRITE operations.

Table~\ref{tab:all_agents_tool_category_pass_fail-combined} presents the distribution of the top seven most frequently invoked categories across all agents, separated by repair outcome. The categories include WRITE (code modification), READ (file content retrieval), TEST (test execution), BUILD (compilation), SEARCH\_CONTENT (pattern search), SEARCH\_FILES (file system queries), and NAVIGATE (directory changes). Each value represents the percentage of commands in that category relative to the agent's total command executions for that outcome class.

The data reveals two primary failure modes: \textit{over-modification without validation} and \textit{over-exploration without action}. \geminicli demonstrates the first mode: WRITE increases from \changeinline{25.8\%} to \changeinline{42.9\%} while TEST decreases from \changeinline{23.2\%} to \changeinline{17.5\%}, yielding a modification-to-comprehension ratio shift from \changeinline{0.97:1} (Pass) to \changeinline{1.99:1} (Fail). \claudecode demonstrates the second mode: WRITE \textit{decreases} from \changeinline{22.6\%} to \changeinline{17.6\%} while SEARCH\_CONTENT \textit{increases} from 8.9\% to \changeinline{14.6\%} in failed repairs, producing a lower modification-to-comprehension ratio (\changeinline{0.74:1}) when failing. \codex maintains consistently minimal comprehension effort (5\% READ) regardless of outcome, yielding an 8:1 modification-to-comprehension ratio in both successful and failed repairs. 
For \codex, BUILD operations account for \changeinline{10.0\%} of all tool commands in failed repairs, compared to \changeinline{6.0\%} in successful repairs, indicating repeated recompilation.
\qwencode allocates \changeinline{16.5\%} to NAVIGATE operations, unique among all agents, suggesting excessive directory traversal overhead.

A common pattern emerges: failed repairs allocate more effort to BUILD operations across \changeinline{two of three} agents (\qwencode: \changeinline{8.4\%}→\changeinline{10.1\%}, \codex: \changeinline{6.0\%}→\changeinline{10.0\%}, \claudecode: \changeinline{11.7\%}→\changeinline{11.6\%}), suggesting iterative debugging without convergence. The absence of a universal Pass-Fail pattern indicates that effective repair strategies are agent-specific: \geminicli requires reducing excessive modification and increasing validation, while \claudecode requires reducing excessive search and increasing code modification.


\begin{figure}[h]
\centering
\includegraphics[width=0.95\textwidth]{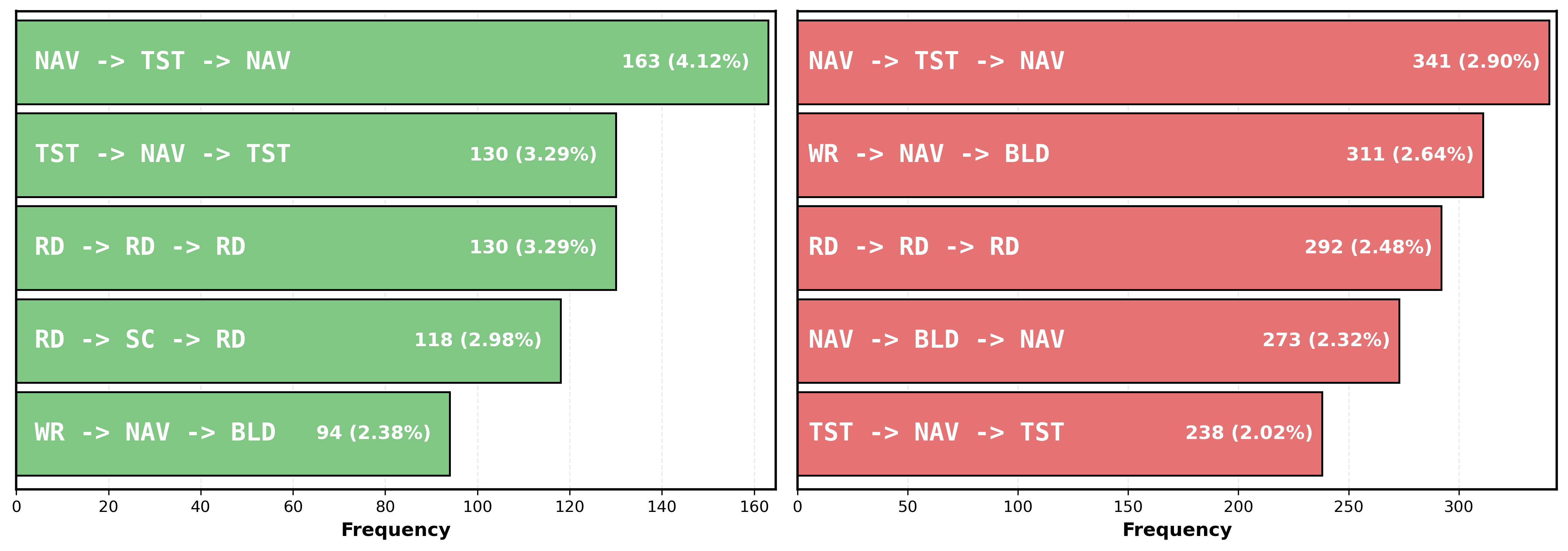}
\caption{Tool sequence patterns for \qwencode.}
\label{fig:qwen_pass_fail_sequences}
\end{figure}


\begin{figure}[t]
\centering
\includegraphics[width=0.95\textwidth]{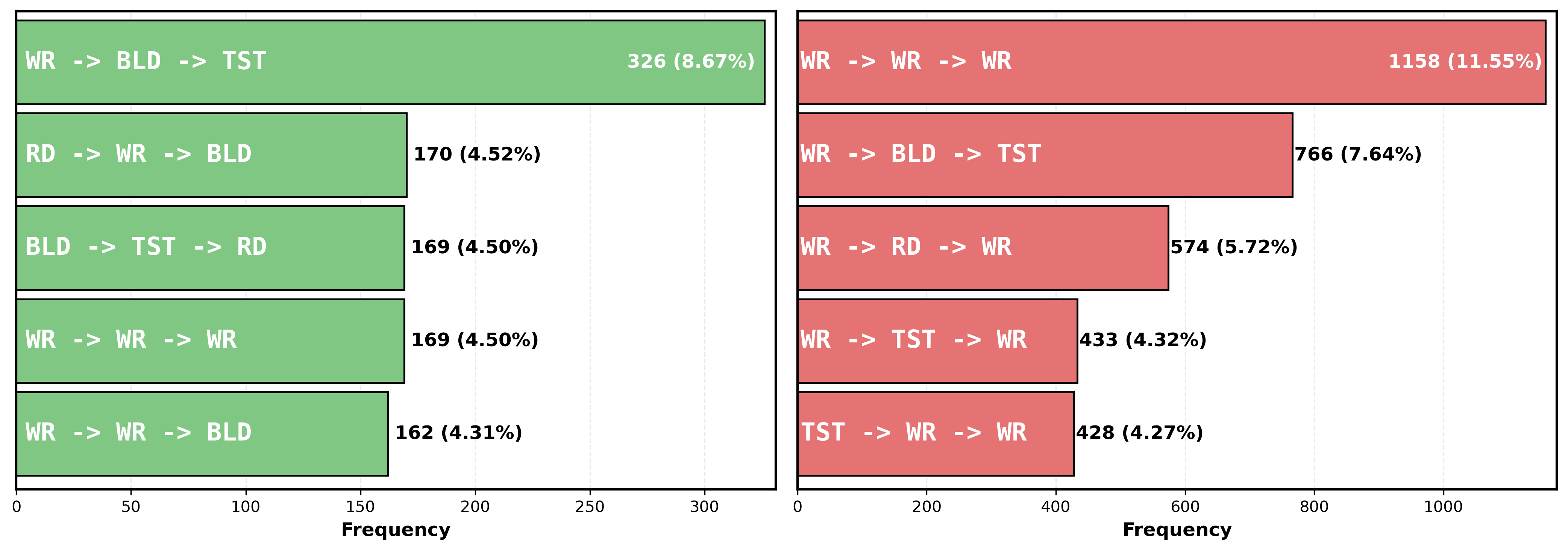}
\caption{Tool sequence patterns for \geminicli.}
\label{fig:gemini_pass_fail_sequences}
\end{figure}

\header{Tool Sequence Patterns in Successful vs Unsuccessful Repairs} 
To understand how repair strategies differ between successful and unsuccessful attempts, we analyze tool sequence patterns using a sliding window approach with $w=3$. A relatively short window ($w=3$) is chosen to capture local action dependencies without diluting interpretability across longer and more variable trajectories. Additional results for window sizes $w=4$ and $w=5$ are included in our replication package.

We extract consecutive 3-step tool invocations from each trajectory and count pattern frequencies for the two lowest-performing agents, \qwencode and \geminicli. Figures~\ref{fig:qwen_pass_fail_sequences} and \ref{fig:gemini_pass_fail_sequences} show the top five patterns separated by repair outcome (abbreviated as WR=WRITE, RD=READ, TST=TEST, BLD=BUILD, NAV=NAVIGATE, SC=SEARCH\_CONTENT). The distribution of tool-use patterns exhibits a long-tail characteristic, with many distinct sequences occurring infrequently. We report both frequency counts and normalized percentages for the most frequently observed patterns to aid interpretability. The relatively low percentages stem from the extensive diversity of distinct action sequences.

\qwencode exhibits navigation overhead in both successful and unsuccessful repairs. The most frequent pattern is NAVIGATE $\rightarrow$ TEST $\rightarrow$ NAVIGATE (\changeinline{4.12\%} successful, \changeinline{2.90\%} unsuccessful). NAVIGATE appears in 3/5 successful patterns and 4/5 unsuccessful patterns. WRITE $\rightarrow$ NAVIGATE $\rightarrow$ BUILD occurs at \changeinline{2.38\%} in successful repairs and \changeinline{2.64\%} in unsuccessful repairs. Navigation overhead persists regardless of outcome.

\geminicli exhibits different patterns in successful versus unsuccessful repairs. Successful repairs use WRITE $\rightarrow$ BUILD $\rightarrow$ TEST most frequently (\changeinline{8.67\%}). Unsuccessful repairs use WRITE $\rightarrow$ WRITE $\rightarrow$ WRITE more frequently (\changeinline{11.55\%}), a \changeinline{2.57-fold} increase over successful repairs (\changeinline{4.50\%})\footnote{This signal compounds in longer windows: four consecutive WRITE operations appear in 8.74\% of unsuccessful four-step windows versus 2.40\% of successful (a 3.64$\times$ gap), and five consecutive WRITE operations in 7.40\% versus 1.38\% of five-step windows (5.36$\times$).}.
WRITE $\rightarrow$ READ $\rightarrow$ WRITE occurs at \changeinline{5.72\%} in unsuccessful repairs. Unsuccessful repairs accumulate consecutive modifications without intermediate compilation or testing.

\qwencode and \geminicli exhibit distinct failure modes. \qwencode shows navigation overhead in both successful and unsuccessful repairs. \geminicli shows a behavioral shift: successful repairs use WRITE $\rightarrow$ BUILD $\rightarrow$ TEST (\changeinline{8.67\%}), while unsuccessful repairs use WRITE $\rightarrow$ WRITE $\rightarrow$ WRITE (\changeinline{11.55\%}). \qwencode requires architectural changes to reduce navigation overhead. \geminicli requires strategies to prevent consecutive modifications without validation.

\subsection{Role of context-assistance with \toolname (RQ6)}

As shown in Table~\ref{tab:multiagent-accuracy-combined}, \qwencode and \geminicli achieve substantially lower repair accuracy than \claudecode and \codex. We examine whether providing repository-level context through \toolname can improve the accuracy of these less effective agents.
\begin{change}
We selected 10 random bugs from each of the five spatial proximity classes in \hunkdefectsforj, yielding 50 bugs. We combined this subset with all 32 \hunkswe bugs, for a total of 82 bugs. \end{change} We ran both \qwencode and \geminicli in this subset under two conditions: baseline (without \toolname) and with \toolname. \toolname provides structured search capabilities through MCP tools (Table~\ref{tab:mcp-tools}) that equip coding agents with intra-repository tools to query class definitions, method implementations, and code snippets across the repository. Table~\ref{tab:multiagent-accuracy-with-maple-combined} presents the impact of \toolname on these agents.


\begin{change}
\header{Impact on Localization} \toolname improves localization for both agents. The localization accuracy of \geminicli increases from 42 bugs (51.22\%) to 48 bugs (58.54\%) with \toolname, a relative improvement of 14.29\%. In contrast, \qwencode increases from 34 bugs (41.46\%) to 35 bugs (42.68\%), a relative improvement of 2.94\%. \end{change} This improvement indicates that \geminicli can effectively leverage structured repository search tools when such tools are available, successfully identifying \changeinline{6} additional fault locations that were previously missed. This finding suggests that context-assistance tools yield benefits primarily for agents with sufficient baseline reasoning capacity to exploit the additional information.


\begin{change}
  \begin{table*}[t]
  \scriptsize
  \centering
  \caption{Impact of \toolname on coding agents for multi-hunk bug repair on \benchmarkdataset}
  \label{tab:multiagent-accuracy-with-maple-combined}
  \begin{tabular}{@{}lcc|cc|cc|cc@{}}
  \toprule
  \textbf{Agent} & \multicolumn{2}{c|}{\textbf{Localization}} & \multicolumn{2}{c|}{\textbf{Compilation}} & \multicolumn{2}{c|}{\textbf{Regression}} & \multicolumn{2}{c}{\textbf{Accuracy}} \\
  & \multicolumn{2}{c|}{\textbf{Success (\%)}} & \multicolumn{2}{c|}{\textbf{Success (\%)}} & \multicolumn{2}{c|}{\textbf{Reduction (Avg)}} & \multicolumn{2}{c}{\textbf{(\%)}} \\
  \cmidrule(lr){2-3} \cmidrule(lr){4-5} \cmidrule(lr){6-7} \cmidrule(lr){8-9}
  & \textbf{Baseline} & \textbf{+\toolname} & \textbf{Baseline} & \textbf{+\toolname} & \textbf{Baseline} & \textbf{+\toolname} & \textbf{Baseline} & \textbf{+\toolname} \\
  \midrule
  \qwencode  & 34 (41.46\%) & 35 (42.68\%) & 77 (93.90\%) & 77 (93.90\%) & -1.30 & -1.62 & 24 (29.27\%) & 26 (31.71\%) \\
  \geminicli & 42 (51.22\%) & 48 (58.54\%) & 81 (98.78\%) & 76 (92.68\%) & -0.96 & -1.78 & 39 (47.56\%) & 47 (57.32\%) \\
  \bottomrule
  \end{tabular}
  \end{table*}
\end{change}

\begin{change}
\header{Impact on Repair Accuracy} The repair accuracy results reveal a promising pattern. \geminicli yields a substantial accuracy gain, rising from 39 bugs (47.56\%) to 47 bugs (57.32\%) with 8 additional correct repairs, representing a 20.51\% relative improvement. With the inclusion of \toolname, \geminicli transitions from fixing fewer than half of the bugs to successfully repairing the majority. \qwencode shows a more modest improvement, increasing from 24 bugs (29.27\%) to 26 bugs (31.71\%).
\end{change}


\begin{change}
These gains come with trade-offs in compilation success and regression metrics. Compilation success remains unchanged for \qwencode at 93.90\%, while it decreases for \geminicli from 98.78\% to 92.68\%. Regression reduction also changes: \qwencode moves from -1.30 to -1.62, and \geminicli moves from -0.96 to -1.78. \end{change} These changes suggest that \toolname encourages more exploratory repair attempts, occasionally leading to unintended side effects. However, the overall repair success for \geminicli, \changeinline{8} additional correct fixes, demonstrates that the accuracy gains outweigh the quality penalties in the aggregate. The trade-offs reflect a fundamental challenge in agentic repair: a richer context enables more ambitious fixes but also increases exploration risk.

\begin{change}
\begin{table}[t]
\scriptsize
\centering
\caption{\toolname{} usage across agents on \benchmarkdataset}
\label{tab:maple-tooluse}
\begin{tabular}{@{}lcc@{}}
\toprule
\textbf{Agent} & \textbf{Fine-grained retrieval} & \textbf{Coarse-grained retrieval only} \\
 & \textbf{(source code)} & \textbf{(\code{repo\_structure})} \\
\midrule
\geminicli & 68/82 (83\%) & 0/82 (0\%) \\
\qwencode  & 45/82 (55\%) & 11/82 (13\%) \\
\bottomrule
\end{tabular}
\end{table}
\end{change}

\begin{change}
\header{Agent Interaction with \toolname{}}
The nine \toolname tools differ in the granularity of the information they return (Table~\ref{tab:mcp-tools}). Eight tools provide fine-grained source code (e.g., a class declaration, method body, code snippet, or class skeleton), while one tool, \code{maple\_repo\_structure}, provides a coarse-grained directory tree.

Table~\ref{tab:maple-tooluse} summarizes, for each agent on \benchmarkdataset, how often fine-grained code retrieval is used in comparison to relying solely on coarse-grained directory information. \geminicli uses fine-grained retrieval for 68 of 82 bugs (83\%) and never relies exclusively on the directory tree. In contrast, \qwencode uses fine-grained retrieval for 45 of 82 bugs (55\%) and relies solely on the directory tree for 11 of 82 bugs (13\%).

The agent that more consistently retrieves fine-grained source code (\geminicli) achieves larger accuracy gains with \toolname compared to the agent that relies less on fine-grained retrieval (\qwencode). This suggests that the effectiveness of \toolname depends critically on retrieving code at the appropriate level of granularity, rather than relying on coarse repository structure alone.
\end{change}

\section{Discussion}
\label{sec:discussion}

Our empirical study of four LLM-driven coding agents on multi-hunk repair tasks reveals critical insights into their capabilities and limitations. The results extend beyond benchmarking coding agents, providing a nuanced view of agent behavior that has implications for practitioners, researchers, and tool builders.

\header{Beyond Accuracy to Trustworthiness}
\label{subsec:trustworthiness}
Our findings demonstrate that repair accuracy is an incomplete metric when considered in isolation. Agents like \qwencode and \geminicli, despite achieving moderate success rates, introduce a substantial ``regression tax" by frequently causing new test failures. This behavior makes them a liability in automated CI/CD pipelines, where stability is crucial. In contrast, high-performing agents like \claudecode and \codex show positive regression reduction, indicating they produce more reliable fixes. Practitioners should pay attention to selecting an agent that prioritizes not only its success rate but also its regression behavior and the high computational cost of its failed repairs. Agents with high-variance outcomes require mandatory human oversight, while those with predictable, high-quality output are closer to enabling autonomous deployment.

\header{Localization Criterion and Patch Equivalence}
\label{subsec:localization-criterion}
Our localization metric operates at the file level and measures whether agents modify the same files as those changed in developer patches. File-level localization provides a clear and reproducible basis for comparing agents, using developer patches as the ground truth. However, it overlooks several nuances that affect interpretation.

Semantically equivalent patches may modify different files. For example, a null pointer exception can be fixed either by adding a null check at the call site or by updating the called method to handle null inputs. Both eliminate the defect but involve different files, which may cause the metric to underestimate localization capability when agents identify valid alternative repair locations. 

Codec\_13 provides a concrete example. The bug presents itself as a potential null pointer exception in \code{DoubleMetaphone.java} when comparing encoding results. The developer patch addresses this through a multi-file refactoring: it removes \code{CharSequenceUtils.java}, adds a null-safe \code{equals()} method to \code{StringUtils.java}, and modifies \code{DoubleMetaphone.java} to use \code{StringUtils.equals()} instead of direct comparison~\footnote{\url{https://github.com/agentic-se/agentic-multihunk-repair/blob/main/Codec_13_patches/Codec_13_ground_truth.diff}}. This approach spans three files. In contrast, \claudecode implements explicit null checking directly within \code{DoubleMetaphone.java}, using intermediate variables and conditional logic to safely handle null values\footnote{\url{https://github.com/agentic-se/agentic-multihunk-repair/blob/main/Codec_13_patches/Codec_13_claude.diff}}. Both patches eliminate the defect and pass all tests. 

These factors help explain cases where localization success appears lower than repair accuracy for high-performing agents such as \codex and \claudecode. These agents may generate correct patches by modifying semantically equivalent file sets or by applying broad edits within a correctly localized file that happen to resolve the defect.

\header{Localization-Repair Gap}
\label{subsec:localizationgap}
A key finding is the \emph{localization–repair gap}, where agents can accurately identify all files requiring modification yet still fail to synthesize a correct patch.
This gap highlights a limitation of current agentic architectures; effective localization does not necessarily translate into coherent reasoning or coordinated editing across interdependent code regions.
However, for complex bugs such as the ones unfixed by all four agents (See Table~\ref{tab:unsolved-bugs-comprehensive}), localization remains a challenge. Metrics such as hunk divergence and spatial proximity are predictors of repair difficulty and can serve as benchmarks for evaluating future progress in this area.

\begin{change}
\header{Qualitative Case Study}
\label{subsec:case-studies}
To explain the localization--repair gap, we analyze three unsolved bugs from Table~\ref{tab:unsolved-bugs-comprehensive}. The bugs represent high, moderate, and low dispersion. These case studies show different failure scenarios across agents.


At high dispersion, we present a case of incomplete edits using \code{JacksonDatabind\_103} (\hunkdefectsforj, Sprawl, 26 hunks across 16 files, div.\ 1.599). The fix requires similar code changes at multiple locations. As part of this fix, it also replaces \code{e.getMessage()} with \code{ClassUtil.exceptionMessage(e)}. All four agents produce partial edits. \claudecode modifies 3 of 16 ground-truth files. \codex modifies 4 files, including 2 correct and 2 incorrect ones. \geminicli modifies a single file with 51 hunks. \qwencode performs many tool calls (86) but produces no valid patch. These results show that agents focus on local context and fail to apply the change to all required locations. They do not generalize the edit across similar sites. This case highlights the need for systematic expansion of edits. When a bug requires repeated changes across multiple locations, agents should first identify the full set of relevant locations before applying edits. A promising direction is to introduce a call-site enumeration step that collects candidate locations through repository search or static analysis, and then verifies during editing that all relevant sites have been addressed.

\begin{change}
At moderate dispersion, three of four agents converge on the same incomplete fix for \code{Astropy\-8707} (\hunkswe, Orbit, 6 hunks across 2 files, divergence 0.450). While \claudecode, \codex, and \geminicli correctly localize the bug, they all apply an incomplete strategy: decoding \code{bytes} at the function entry and leaving the rest of the code unchanged. The ground truth patch updates internal comparison sites to ensure consistent handling of \code{bytes} and \code{str}. In contrast, the agents do not propagate the change beyond the entry point. As a result, the original test still fails, and additional regressions are introduced. The key insight is that these agents apply \emph{entry-only fixes}: they adjust the input but do not follow how that change affects the rest of the function. A practical takeaway is that after modifying a variable (e.g., decoding the input), agents should trace where that variable is used within the function—such as in comparisons, loops, or conditional checks—and verify that each use still behaves correctly. This requires understanding the modified function and systematically checking all dependent code before finalizing the patch.
\end{change}

Another high-dispersion example shows how partial multi-file coordination can introduce new failures. \code{Closure\_157} (\hunkdefectsforj, Sprawl, 7 hunks across 3 files, divergence 1.078) requires coordinated changes, and none of the agents modify all three files. \claudecode and \codex each modify two files and successfully fix most failing tests (12/12 and 10/12, respectively). However, both introduce two new failures in \code{RenamePrototypesTest}, which correspond to the unmodified third file. The root cause is that the modified files change the internal representation of object-literal keys, while the unmodified file still assumes the old representation. The key insight is that partial fixes across related files can break consistency: changing one part of a codebase without updating its dependents may introduce new errors. A practical takeaway is that when new test failures appear in files the agent did not modify, they should be treated as contextual cues of missing edits. Agents should then trace which components depend on the changed behavior and ensure that all affected files are updated before finalizing the patch.

\end{change}

\header{Agent Efficiency}
\label{subsec:efficiency}
Failed repairs impose substantial computational costs, as coding agents often continue to consume resources rather than terminate early. For practitioners evaluating production-ready systems, this highlights the importance of assessing not only repair accuracy but also the efficiency of failure handling. High-performing agents such as \claudecode and \codex justify their use of resources through consistently higher success rates, whereas low-performing agents incur compounded costs—low success combined with high resource expenditure per failure, making them less suitable for large-scale deployment.

These findings point to concrete directions for tool improvement. For agentic tool developers, our results emphasize that consistency and semantic understanding are as critical as raw coding ability. Agents should be engineered to detect unproductive repair paths and “fail fast,” conserving computational resources for promising attempts. The effectiveness of \claudecode’s caching mechanism further illustrates the value of efficient context management in reducing redundant computation and improving overall repair efficiency.

  \header{Implications for Context-Driven Repair}
  \toolname raises the repair accuracy of \geminicli but has little effect on \qwencode, and the difference comes from how each agent uses the tools. \begin{change} For example, \code{Gson\_9} is repaired by adding a \code{value(Boolean)} overload to the \code{JsonWriter} class. \geminicli called \code{maple\_find\_class} on \code{JsonWriter}, called \code{maple\_find\_method\_in\_class} on its \code{value} method, added the overload, and called \code{maple\_find\_method\_in\_file} to check the change against the failing test; the repair passed. For the same defect, \qwencode called \code{maple\_repo\_structure} to list the directory, then called \code{maple\_find\_class} on \code{JsonPrimitive}, a class that the fix does not modify, and produced an empty patch. Both agents had access to the same nine \toolname tools, yet only \geminicli directed its queries at the class that the fix modifies. \end{change} The benefit of context assistance, therefore, depends on whether the agent queries the code that the fix must modify, and not on the general reasoning ability of the agent. A direct next step is to guide the agent toward the classes and methods named in the failing test before it edits, so that the retrieved context covers the location of the fix.

\header{Non-Deterministic Nature of Runs}
\label{subsec:nondeterminism}
Agentic multi-hunk repair is inherently non-deterministic. Even when provided with identical inputs, repeated runs of the same agent may follow distinct trajectories and yield different outcomes. This variability stems from the probabilistic nature of the underlying LLM and the continuous feedback-driven interaction between the agent and the development environment.

During repair, even a minor variation in the internal reasoning of the model, such as proposing an alternative fault hypothesis, selecting a different file to edit, or generating a slightly different patch, can alter the trajectory of the agent. The agent continuously updates its plan based on compiler diagnostics, test outcomes, and observed code changes, and each reasoning step directly affects the subsequent actions. This feedback loop introduces path dependency, where small perturbations in reasoning responses can propagate into substantially different sequences of actions.

These non-deterministic effects are particularly amplified in multi-hunk repair. Coordinated modifications across multiple interdependent code regions require consistent reasoning over distributed contexts. Small deviations, such as addressing one hunk prematurely or misinterpreting a diagnostic message, can disrupt this coordination and lead to incoherent or incomplete patches.

\header{Auto-Update Behavior of the Agents}
\label{subsec:autoupdate}
A controlled evaluation of coding agents requires stable and reproducible environments. During our experiments, we observed unexpected instances of automatic software updates initiated by both the \qwencode and \claudecode agents. In the case of \qwencode, the agent initiated a self-update without explicit user permission while executing a repair task, displaying the message \code{Update successful! The new version will be used on your next run.}

For \qwencode, the experiment began with version~0.0.11, which automatically upgraded to version~0.0.14 during execution, as verified using the command \code{qwen --version}. Similar update behavior, though without explicit notification, was observed for \claudecode.





\header{Operational Instability of Models}
\label{subsec:modeloutage} The execution of agentic systems depends on the continuous availability and stability of their hosted model infrastructure. During our evaluation, we observed several external factors that affected the operational consistency of \claudecode and other hosted agents.\footnote{\url{https://status.claude.com/}} The first factor involves service outages, where the Claude status dashboard reported elevated error rates for the Sonnet~4.5 model and temporary unavailability of the Claude API and \claudecode interfaces.\footnote{\url{https://status.claude.com/incidents/ggjp05h790b3}} A second factor concerns degraded service quality, caused by an upstream provider error that exposed a fault in Anthropic infrastructure, leading to intermittent access issues and increased latency across the agent and API.\footnote{\url{https://status.claude.com/incidents/gr1vrcvz9jd4}} A third factor relates to dynamic service recovery, where fluctuating system states during mitigation intervals resulted in inconsistent agent responses and incomplete repair trajectories.

These operational variations highlight the dependency of coding agents on third-party hosted environments. Even short-lived degradations or transient failures can interrupt long-running repair sessions, influence success rates, and reduce experimental reproducibility. Establishing controlled execution environments and reliable service baselines remains essential for rigorous empirical evaluation of agentic systems.










\header{Human-in-the-loop} 
\label{subsec:human-in-the-loop}
In this work, we evaluate agents in a fully autonomous mode. In reality, these are used as assistants to human developers. A failed patch might still provide a valuable starting point for a human to fix the remaining issues. Future work could explore how agents can serve as assistants, generating partial or candidate fixes for complex bugs, with a human developer guiding, validating, and completing the repair. This would also necessitate developing new metrics, such as the `closeness to the ideal fix', to quantify the utility of imperfect patches.

\begin{change}
\header{Cross-Language Behavior of Coding Agents} \benchmarkdataset combines multi-hunk bugs from Java and Python, and the two subsets differ in the information provided for each bug. \hunkdefectsforj bugs are drawn from Defects4J~\cite{defects4j} and include a short bug report. \hunkswe bugs are drawn from SWE-bench Verified as GitHub issues, which often provide richer context and may reference relevant files or functions. These differences should be considered when interpreting cross-language results.

Repair accuracy is higher on Python than on Java for three of the four agents. The largest gap is observed for \geminicli, which achieves 41.67\% on Java and 59.38\% on Python. Only \claudecode attains higher repair accuracy on Java (93.28\%) than on Python (87.50\%).

Localization success is also higher on Python for all four agents, ranging from 65.63\% for \qwencode to 90.63\% for \claudecode. In contrast, regression reduction is lower on Python for all agents. \claudecode decreases from +2.47 on Java to -1.43 on Python, while \codex drops from +2.25 to +1.16. This pattern is particularly pronounced for \claudecode, which exhibits net-positive regression behavior on Java but net-negative behavior on Python. In contrast, \codex maintains net-positive regression reduction on both languages.

Repair accuracy and regression behavior do not always align, and their relationship varies across languages. These results suggest that evaluations of coding agents should account for programming language in addition to accuracy and regression metrics. Aggregate scores computed over a single-language benchmark may obscure language-specific differences in agent behavior. Future evaluation frameworks should therefore consider results per language, enabling practitioners to select agents that match the language characteristics of their target codebases.
\end{change}

\begin{change}
\header{Multilingual Generalization of Hunk Divergence and Proximity} The hunk divergence and spatial proximity metrics were originally defined and validated on the Java instances of \hunkdefectsforj~\cite{birch}. In this work, we extended these metrics to a different language (Python) and applied them to \hunkswe. The findings on Java and Python are the same. Fixed bugs exhibit lower divergence than unfixed bugs. For \geminicli, the median divergence rises from 0.38 to 0.47 on Java and from 0.12 to 0.31 on Python. For \claudecode, it rises from 0.41 to 0.58 on Java and from 0.16 to 0.24 on Python. Repair accuracy also declines on average as bugs become more dispersed across the codebase. For \codex on Java, accuracy ranges from 91.53\% on Nucleus bugs to 63.64\% on Fragment bugs. These patterns hold across all four agents evaluated. The replication suggests that edit heterogeneity and dispersion are properties of multi-hunk repair rather than of any single language. The current evidence covers two languages and four agents; broader claims would require evaluation on additional languages and agents.
\end{change}

\begin{change}
\header{\toolname Generalizes Across Languages}
The \toolname results in Table~\ref{tab:multiagent-accuracy-with-maple-combined} aggregate Java and Python bugs. A per-language breakdown, reported in the repository~\cite{repo}, shows that \toolname improves repair accuracy in both languages. For \geminicli, accuracy increases from 40.00\% to 52.00\% on Java and from 59.38\% to 65.63\% on Python. For \qwencode, accuracy increases from 22.00\% to 24.00\% on Java and from 40.63\% to 43.75\% on Python. These results indicate that the benefit of repository-level context is not specific to a single language ecosystem. Therefore, our tool, \toolname, an MCP-based context-assistance tool, generalizes across languages.
\end{change}

\header{Towards Divergence-Guided Agent Selection} Multi-hunk bugs in \benchmarkdataset exhibit substantial variation in structural and spatial divergence, and no single coding agent consistently demonstrates high accuracy across this spectrum, as shown in Figure~\ref{fig:venn-llm} and Table~\ref{tab:agent-performance-combined}. \claudecode contributes 25 unique repairs that no other agent produces, \codex contributes 12, and \geminicli contributes 3, indicating that even the strongest agents do not subsume one another. The per-language results in Table~\ref{tab:multiagent-accuracy-combined} reinforce this point: \claudecode attains the highest repair accuracy on Java (93.28\% on \hunkdefectsforj), whereas \codex attains the highest accuracy on Python (90.63\% on \hunkswe). We advocate for \emph{divergence-aware agent selection} that estimates fix dispersion using pre-repair signals, such as the number of fault-localized regions, their distribution across files and methods, and the breadth of the failing stack trace.

A baseline alternative is post hoc ensembling across all available agents. While conceptually straightforward, this approach is computationally inefficient and incurs substantial inference cost. In our evaluation setup, ensembling across the four evaluated agents on the 404 \benchmarkdataset bugs requires 1{,}616 repair trajectories, even though each agent resolves only a partial subset of bugs, and the subsets overlap heavily. In contrast, a divergence-guided selection mechanism can perform predictive agent selection based on pre-repair signals that estimate the anticipated divergence of the required fix. For instance, bugs that span multiple files and exhibit high hunk divergence can be assigned to agents that maintain strong performance in dispersed settings, such as \claudecode and \codex, both of which exceed 70\% repair accuracy on Fragment-class bugs in \benchmarkdataset. In contrast, lower-divergence bugs confined to a single method or file may be handled by lighter-weight agents at reduced operational cost. A divergence-guided selection mechanism may further exploit per-language strengths by assigning Java and Python bugs differently when the project language is known. Aligning agent selection with the predicted divergence of the bug, as well as with the host language, can improve both computational efficiency and overall repair coverage compared with single-agent or naive ensemble strategies.

\header{Beyond Patch Repair: Divergence-Aware Issue Resolution and Reproduction}  
Our empirical results show that \emph{hunk divergence} and \emph{spatial proximity}, introduced to quantify intra-patch complexity, are correlated with agentic repair success (Table~\ref{tab:agent-performance-combined}). Across all four agents, fixed bugs exhibit lower median divergence than unfixed ones, and repair accuracy declines as bugs become more spatially dispersed. For \claudecode, accuracy drops from 100\% on Nucleus-class bugs to 86.67\% on Fragment-class bugs; analogous declines are observed for \codex (92.31\% to 73.33\%), \geminicli (49.23\% to 33.33\%), and \qwencode (24.62\% to 13.33\%). These results indicate that coding agents struggle increasingly as the spatial and lexical dispersion between hunks grows.  

While our study focuses on program repair, the properties captured by hunk divergence and spatial proximity may extend to other tasks that require reasoning over distributed software artifacts. For instance, GitHub issues often describe failures spanning multiple modules, with stack traces in one component, reproduction steps in another, and implicit dependencies across configuration files. Although we do not evaluate divergence in this setting, our findings suggest that divergence-based metrics could be adapted to issue triaging~\cite{swebench, swe-agent, autocoderover, tao:magis:arxiv24, openhands, agentless, specrover, liu:mars-code-agent:arxiv24} or bug reproduction~\cite{mundler:swt-bench:arxiv25, cheng:agenticbugreproduction:arxiv25, issue2test}. In particular, estimating divergence from the issue reports, such as the number and location of referenced files, could help prioritize complex issues for dynamic analysis or guided reproduction.

\header{Understanding Patch Equivalence and Test Overfitting}
Agentic repair shows that coding agents frequently generate patches that pass all tests but differ structurally from developer patches. It remains unclear whether these patches are semantically equivalent to developer fixes or merely overfit to the test suite while leaving bugs partially unresolved.

Large-scale empirical studies are needed to characterize when and how agent patches diverge from developer patches, to identify structural patterns that distinguish equivalent patches from overfitting ones, and to determine which bug types are most susceptible to test overfitting. Building on these insights, techniques can be developed to detect potential overfitting using program analysis to identify behavioral divergences and targeted test generation to expose incomplete fixes.
\section{Threats to Validity}

\header{Construct Validity} This concerns the relationship between the theoretical concepts being studied and the specific metrics used to measure them. Our defined metrics, namely file localization success, compilation success, regression reduction, and repair accuracy, are designed to capture key aspects of the repair process. However, they are simplifications of a complex reality. For example, our repair accuracy metric is binary (a patch either passes all tests or does not), which does not account for partially correct patches that might still be useful to a developer. Similarly, localization success operates at the file level, meaning an agent could modify the correct files but still fail to address the correct buggy hunks within them. Nevertheless, these metrics provide a clear, objective, and reproducible benchmark for comparing coding agents in the core task of fully automated repair. To provide a more nuanced view, we complemented these outcome-based metrics with a qualitative analysis of tool sequence patterns, which reveals the underlying behaviors and strategies of the agents, offering insights beyond binary success or failure.

\begin{change}
We employ hunk divergence to characterize multi-hunk complexity, following prior work~\cite{birch}. This metric does not capture semantic coordination between hunks, such as dataflow relationships, call-graph proximity, or type-level coupling. Developing semantic measures of multi-hunk complexity is therefore a promising direction for future work. Nevertheless, hunk divergence based on lexical, AST-level, and file-level dissimilarity provides a programming-language-agnostic characterization of multi-hunk complexity across the Java and Python subsets of \benchmarkdataset, without requiring language-specific program analysis.
\end{change}

\begin{change}
We characterize agent behavior by abstracting each tool invocation into a functional action category based on its operational purpose, such as read, write, test, build, and search. This abstraction makes agent-specific native tools and shell commands comparable. A limitation of this action model is that it does not capture contextual attributes of individual actions, such as the file path, structural scope, or file type on which an action operates. For example, an edit may occur within a function, a method, or a module, but the action model does not distinguish among these scopes. While we measure localization and spatial proximity, these metrics characterize the overall repair rather than the individual actions that produce it. Enriching the action model with such contextual attributes would enable analysis of how the spatial focus of an agent evolves over the course of a repair, and we leave this to future work.
\end{change}

\header{Internal Validity}
This refers to the confidence that the observed outcomes are caused by the experimental variables and not by confounding factors.
As discussed in Section~\ref{subsec:nondeterminism}, LLM-based Agentic systems are inherently non-deterministic. The same agent, given the same initial prompt, can produce different action trajectories and final patches across multiple runs due to the probabilistic nature of the underlying LLM. This means a single successful or failed run may not be representative of the agent's true capability. While eliminating this non-determinism is currently impossible, we mitigated its impact by evaluating across a wide range of problems (i.e., on \changeinline{404} distinct multi-hunk bugs from the \changeinline{\benchmarkdataset} dataset). This allows us to observe overarching trends and patterns in agent behavior that are less susceptible to the randomness of a single execution.

As discussed in Section~\ref{subsec:autoupdate}, agents can be affected by external factors such as API outages, service degradation, and automatic background updates. These events are outside of our control and could have impacted the outcomes of specific repair attempts. We addressed this by using a standardized evaluation harness that provided identical access to tools and environments for all agents. We logged all interactions and, where possible, noted the agent versions before and after any auto-updates. While these factors represent a real-world challenge of using agentic systems, our reporting on these instabilities contributes to a more transparent understanding of the current state of these tools.

\header{External Validity} This concerns the generalizability of our findings to other contexts, such as different programming languages, projects, bug types, or coding agents. We evaluated four prominent agentic coding tools: \codex, \geminicli, \qwencode, and \claudecode. However, the field of AI is rapidly evolving, and new, more capable agents may have emerged since our evaluation was conducted. The reported performance numbers may not accurately reflect the capabilities of the latest models. The agents selected were state-of-the-art and widely used at the time of our study, to the best of our knowledge. More importantly, our primary contribution is not just the performance benchmark itself, but the conceptual framework, the behavioral metrics, and the qualitative insights into the challenges of multi-hunk repair. These findings, such as the difficulty in coordinating edits across dispersed locations, are fundamental to the problem and are likely to remain relevant for future generations of agents. Our evaluation methodology can be readily applied to benchmark new tools as they become available.

Our study relies exclusively on the \changeinline{\benchmarkdataset} dataset, which contains multi-hunk bugs from \changeinline{Java and Python projects}. Therefore, our findings may not generalize directly to other programming languages \changeinline{(e.g., C++, JavaScript)} or to bug types not represented in this dataset (e.g., concurrency or performance bugs). However, \changeinline{\benchmarkdataset} is, to our knowledge, the largest and most systematically curated dataset specifically designed for multi-hunk bug repair research, making it the most appropriate choice for this study.
\changeinline{Future work should replicate these findings in other languages and across different project domains to assess the generalizability of the results.}

\begin{change}
The two subsets of \benchmarkdataset differ substantially in size: 372 Java bugs from \hunkdefectsforj and 32 Python bugs from \hunkswe. Consequently, results on \hunkswe may be sensitive to the limited sample size. Expanding the Python subset would further strengthen the generality of our claims.
\end{change}
\section{Related Work}

Autonomous agents iteratively reason, invoke tools, and adapt based on environmental feedback \cite{roychoudhury:agenticai:arxiv25}. Bug repair agents include RepairAgent \cite{bouzenia:repairagent:icse25}, AutoCodeRover \cite{autocoderover}, OpenHands \cite{wang:openhands:iclr24}, SWE-Agent \cite{swe-agent}, Agentless \cite{agentless}, Magis \cite{tao:magis:arxiv24}, AgentCoder \cite{huang:agentcoder:arxiv24}, MarsCode Agent \cite{liu:marscodeagent:arxiv24}, FixAgent \cite{lee:fixagent:arxiv24}, and Passerine \cite{rondon:passerine:arxiv25}. Extensions such as SpecRover \cite{specrover} employ dedicated reproduction and testing phases. These agents employ diverse architectures: iterative reasoning and testing, structured search-locate-fix workflows, and specialized agent-computer interfaces. There are also specialized agents that handle project setup \cite{hu:dockeragent:arxiv25}, test execution \cite{bouzenia:llmtestexec:issta25}, notebook debugging \cite{grotov:debugnotebook:arxiv24}, package installation \cite{milliken:pipinstall:arxiv24}, and log analysis \cite{roy:rootcause:fse24}.

While these agents are employed on diverse software engineering tasks, understanding their internal decision-making processes, operational dynamics, and failure modes remains largely unexplored. Recent work~\cite{bouzenia:understanding:ase25} analyzes trajectories of three agents (RepairAgent, AutoCodeRover, OpenHands) to identify debugging anti-patterns through qualitative analysis. However, their study differs from ours in several fundamental ways. First, they evaluate research prototypes while we evaluate production coding agents used by actual developers (\claudecode, \codex, \geminicli, \qwencode) which leverage the latest SOTA models (\sonnet, \gpt, \geminimodel, \qwencodermodel). Second, they sample 120 trajectories, whereas we conduct a large-scale study on \changeinline{404} multi-hunk bugs (\changeinline{1,616} repair trajectories). Third, they do not systematically characterize multi-hunk bugs.  We leverage hunk divergence and spatial proximity metrics~\cite{birch} to quantify multi-hunk complexity. Fourth, their analysis focuses on qualitative action patterns, while we provide a multi-dimensional quantitative evaluation. Finally, we propose and evaluate Maple, an MCP-based context-assistance mechanism that improves mid-tier agent accuracy by 30\%. The BIRCH benchmark~\cite{birch} specifically targets multi-hunk bugs and introduces the hunk divergence and spatial proximity metrics we employ. 

\begin{change}
Our prior work~\cite{birch, nashid_thesis_2026} introduced the hunk divergence and spatial proximity metrics and evaluated six LLMs on \hunkdefectsforj{} under a limited feedback loop. In contrast, this work makes the following contributions: (a) we study four production-grade coding agents—\claudecode, \codex, \geminicli, and \qwencode—that operate autonomously; (b) we extend the evaluation dataset from the Java-only \hunkdefectsforj{} (372 bugs) to the multi-language \benchmarkdataset{} (404 bugs), incorporating 32 Python bugs from SWE-bench Verified (\hunkswe); (c) we introduce a fine-grained set of metrics to characterize agent behavior beyond repair accuracy, including localization effectiveness, compilation success, regression incidence, and operational dynamics such as token consumption, execution time, and tool utilization; (d) we analyze 1,616 repair trajectories for multi-hunk repair across Java and Python; and (e) we propose \toolname, an MCP-based tool, and investigate the role of intra-repository context in multi-hunk repair by coding agents.
\end{change}

\section{Conclusion}

This work presented the first systematic study of LLM-driven coding agents on the complex task of multi-hunk program repair. We evaluated four agents (\claudecode, \codex, \geminicli, and \qwencode) on \changeinline{404} multi-hunk bugs from the \changeinline{\benchmarkdataset} dataset, introducing fine-grained metrics for localization, repair accuracy, regression behavior, and operational dynamics. We also developed \toolname, an MCP-based tool that improved the repair accuracy of less capable agents.
Our analysis revealed substantial performance variation, with repair accuracy ranging from \changeinline{26.98\%} (\qwencode) to \changeinline{92.82\%} (\claudecode). Success was strongly influenced by bug complexity, as repair accuracy consistently declined sharply with increasing bug dispersion (hunk divergence). 
We identified a distinct localization-repair gap; \changeinline{58.33\%} of the \changeinline{12} bugs unfixed by any agent were successfully localized by at least one agent, yet a correct fix could not be synthesized.
The operational assessment revealed a stark divide in effectiveness based on regression reduction. High-performing agents, \claudecode and \codex, achieved positive average regression reduction, demonstrating superior semantic consistency. 
Conversely, \qwencode (\changeinline{-1.50}) and \geminicli (\changeinline{-2.10}) exhibited negative regression reduction, implying they introduced new test failures, thus imposing a ``regression tax". 
Furthermore, agents do not ``fail fast"; failed repairs are expensive, often consuming significantly more resources (tokens) than successful ones (e.g., \geminicli failed repairs consumed \changeinline{440.4\%} more input tokens).
\begin{change} 
These results highlight not only the accuracy of coding agents but also how they operate, emphasizing the need to move beyond simple pass/fail metrics. Evaluations should instead account for operational behavior, resource efficiency, and the ability to minimize regressions when building production-ready agentic tools.
\end{change}

\section{Data Availability}
The replication package~\cite{repo} includes an automated framework for assessing four coding agents on \changeinline{\benchmarkdataset}, the implementation of \toolname, and instructions for reproducing our experiments.

\bibliographystyle{ACM-Reference-Format}
\bibliography{references}


\end{document}